\documentclass[apj]{emulateapj}

\shorttitle{Formation of a Brightest Cluster Galaxy}

\shortauthors{Rasmussen et al. }

\begin{document}

\title{Witnessing the Formation of a Brightest Cluster Galaxy in a
Nearby X-ray Cluster$^\star$}

\author{Jesper Rasmussen,\altaffilmark{1,4}
  John S.~Mulchaey,\altaffilmark{1} Lei~Bai,\altaffilmark{1}
  Trevor J.~Ponman,\altaffilmark{2} Somak Raychaudhury,\altaffilmark{2}
  and Ali~Dariush\altaffilmark{3}}

\altaffiltext{1}{Carnegie Observatories, 813 Santa Barbara Street,
  Pasadena, CA 91101, USA; jr@obs.carnegiescience.edu}

\altaffiltext{2}{School of Physics and Astronomy, University of
  Birmingham, Edgbaston, Birmingham B15 2TT, UK}

\altaffiltext{3}{School of Physics and Astronomy, Cardiff University,
  Queens Buildings, The Parade, Cardiff, CF24 3AA, UK}

\altaffiltext{4}{Chandra Fellow}

\altaffiltext{$\star$}{This paper includes data gathered with the 6.5 meter
Magellan Telescopes located at Las Campanas Observatory, Chile.}

\begin{abstract}

The central dominant galaxies in galaxy clusters constitute the most
massive and luminous galaxies in the Universe. Despite this, the
formation of these brightest cluster galaxies (BCGs) and the impact of
this on the surrounding cluster environment remain poorly understood.
Here we present multi-wavelength observations of the nearby poor X-ray
cluster MZ\,10451, in which both processes can be studied in
unprecedented detail. {\em Chandra} observations of the intracluster
medium (ICM) in the cluster core, which harbors two optically bright
early-type galaxies in the process of merging, show that the system
has retained a cool core and a central metal excess. This suggests
that any merger--induced ICM heating and mixing remain modest at this
stage. Tidally stripped stars seen around either galaxy likely
represent an emerging intracluster light component, and the central
ICM abundance enhancement may have a prominent contribution from {\em
in situ} enrichment provided by these stars. The smaller of the
merging galaxies shows evidence for having retained a hot gas halo,
along with tentative evidence for some obscured star formation,
suggesting that not all BCG major mergers at low redshift are
completely dissipationless. Both galaxies are slightly offset from the
peak of the ICM emission, with all three lying on an axis that roughly
coincides with the large-scale elongation of the ICM.  Our data are
consistent with a picture in which central BCGs are built up by
mergers close to the cluster core, by galaxies infalling on radial
orbits aligned with the cosmological filaments feeding the cluster.

\end{abstract}

\keywords{galaxies: clusters: individual (MZ\,10451) --- galaxies:
elliptical and lenticular, cD --- galaxies: evolution --- galaxies:
interactions --- X-rays: galaxies: clusters}

\section{INTRODUCTION}\label{sec,intro}

The most massive galaxies in the Universe occur in rich groups and
clusters. These brightest cluster galaxies (hereafter BCGs) often
reside close to the peak of the diffuse X-ray emission from the
intracluster medium (ICM; \citealt{jone84,zabl98,lin04,raff08}), and
can have recessional velocities indistinguishable from the cluster
mean (\citealt{quin82,zabl90}; but see also \citealt{cozi09}). Their
major axis is preferentially aligned with that of the host cluster, as
defined by the projected distribution of cluster galaxies (e.g.,
\citealt{bing82}) or the ICM X-ray emission \citep{hash08}, an effect
which seems independent of cluster richness \citep{full99}.
Furthermore, there is a clear relationship between the stellar
luminosity of central BCGs and the total mass and X-ray luminosity of
their host cluster \citep{edge91,brou08,hans09,mitt09}.

These observations suggest that many BCGs reside near the centers of
the cluster gravitational potential, and that the formation of central
BCGs is intimately linked to the formation and evolution of the
cluster itself, as anticipated in a hierarchical structure formation
scenario. This idea is supported by results of cosmological
simulations which indicate that central BCGs are built up through a
series of mergers near the cluster core (e.g,
\citealt{dubi98,conr07,delu07}). Such simulations also suggest that
these mergers must have been predominantly dissipationless, at least
for BCG major mergers occurring at low redshift (e.g.,
\citealt{khoc03,delu07}). The notion that BCGs have a unique formation
history is further reinforced by findings indicating that their
properties are separate from those of other (cluster) galaxies. For
example, the optical luminosity of BCGs is not drawn from the same
distribution as that of other cluster members \citep{trem77,vale08},
BCGs tend to have spatially extended stellar halos not seen around
other galaxies (e.g., \citealt{gonz03}), and they are more likely to
be radio-loud than other galaxies of similar stellar mass
\citep{best07}.

Unfortunately, an observational understanding of the formation of BCGs
is largely limited by the fact that few systems in the process of
forming have actually been identified. Multiple nuclei are common in
BCGs at low and moderate redshifts, suggesting that galaxy--galaxy
mergers must be frequent (e.g., \citealt{laue88, mulc06, jelt07}). In
a few cases, multiple galaxies at the initial stages of merging have
been identified at the centers of X-ray clusters
\citep{yama02,nipo03,tran08}, including one case in which tidal
features provide direct evidence for galaxy--galaxy interactions
\citep{rine07}. However, as these examples all occur at moderate to
high redshifts, detailed studies of the BCG formation process have not
been possible. Potential examples at lower redshift include known
dumbbell galaxies \citep{greg92}, some of which constitute the
brightest galaxy in X-ray clusters. However, at least some of these
are likely to represent various stages in the mergers of {\em cluster}
cores, each already containing a fully formed central dominant galaxy
\citep{trem90,pimb08}. As such, they may not offer a particularly
clean view of the build-up of the BCG itself.

For these reasons, it also remains uncertain to what extent BCG
formation may have affected the thermodynamic history of baryons in
X-ray bright cluster cores. For example, major galaxy--galaxy mergers
near the core of a cluster could provide a mechanism for distributing
metals into the ICM \citep{zari04} and so provide a partial
explanation for the pronounced central metal excesses observed in many
systems (e.g., \citealt{fino00,degr04,lecc08}). Such mergers might
also act to disrupt the formation of cool cores, and thereby help
explain the puzzling dichotomy between ``cool-core'' and ``non
cool-core'' groups and clusters (e.g., \citealt{sand09}).

Here we present a {\em Chandra} study of a recently discovered
low-redshift ($z\simeq 0.06$) X-ray cluster, MZ\,10451, with a BCG in
the process of forming. To the best of our knowledge, and if excluding
dumbbell galaxies but including major merger candidates in
low-redshift SDSS groups and clusters \citep{mcin08}, this represents
the first known case of an ongoing central BCG major merger in an
X-ray cluster in the nearby Universe. The goal of the present
investigation is two-fold. We aim to establish how the ongoing
galaxy--galaxy interaction is affecting the thermodynamics and
chemical properties of the ICM in the cluster core. The relatively low
ICM temperature of this system, $T \approx 1$~keV, makes it attractive
for a study of this kind, because any X-ray signatures of the
interaction are likely to be more pronounced than in much more massive
clusters. Secondly, we also hope to shed light on the BCG formation
process itself, e.g.\ by exploring whether the merging galaxies
contain any hot (or cold) gas (i.e.\ whether the merger is
dissipationless), and whether they show evidence for strong nuclear
activity induced by the interaction.

We assume $H_0=73$~km~s$^{-1}$~Mpc$^{-1}$, $\Omega_m=0.27$, and
$\Omega_\Lambda=0.73$. The target redshift of $z=0.0607$ then
corresponds to a luminosity distance of $D \approx 260$~Mpc, and $1'$
on the sky to 67~kpc. We further adopt Solar abundances from
\cite{ande89}, and a local Galactic H{\sc i} absorbing column of
$N_{\rm H}=1.7\times 10^{20}$~cm$^{-2}$ \citep{dick90}. Unless
otherwise stated, uncertainties are given at the 68\% confidence
level.

\section{MZ\,10451: A BCG IN FORMATION}

As part of the {\em XI} Groups Survey \citep{rasm06}, we have obtained
comprehensive multi-wavelength data for a sample of 25 groups and poor
clusters. We are using {\em XMM-Newton} to study the X-ray properties
of each system, and the IMACS multi-object spectrograph \citep{dres06}
on the 6.5-m Baade/Magellan telescope at Las Campanas to determine
group membership down to very faint luminosities ($M_R \approx
-15$). In addition, we have obtained {\em Spitzer}/MIPS and {\em
GALEX} UV imaging of each group field to probe the star formation
properties of the group members. Details of the X-ray, optical, and
infrared data reduction can be found in \citet{rasm06} and
\citet{bai10}, while the {\em GALEX} analysis will be described in a
forthcoming paper (J.~Rasmussen et~al., in preparation). A summary of
the X-ray, UV, and $24\,\mu$m data considered in the present paper is
provided in Table~\ref{tab,log}.

\begin{table}
\caption{Summary of MZ\,10451 observations discussed in this paper\label{tab,log}}
\begin{center}
\begin{tabular}{lccc}
  \tableline \hline
Instrument  & Obs.\ ID & Obs.\ Date & $t_{\rm expo}$ \\
           &            & (yyyy-mm-dd) &  \\ \hline
{\em Chandra} ACIS-S  & 10467      &  2008-11-14 & 18.78~ks \\
{\em Chandra} ACIS-S  & 10800      &  2008-11-04 & 28.16~ks \\
{\em Chandra} ACIS-S  & 10802      &  2008-11-16 & 31.62~ks \\
{\em XMM} pn      & 0305800901 &  2006-01-20  & 14.99~ks \\
{\em XMM} MOS1    & \ldots     & \ldots       & 21.17~ks \\
{\em XMM} MOS2    & \ldots     & \ldots       & 21.21~ks \\
{\em GALEX}          &   GI4\_052009         &   2008-09-27   & 1607~s \\ 
{\em Spitzer} MIPS   &   22243840     &  2007-08-23  & 84~s\\
  \tableline
\end{tabular}
\end{center}
\end{table}

One of the most X-ray luminous systems in the {\em XI} Survey so far
is MZ\,10451. Our optical spectroscopy of this system has identified
34 group members within the $r=15'$ IMACS field-of-view, with a
velocity dispersion $\sigma_v \approx 500$~km~s$^{-1}$. A search of
the 2dF galaxy redshift catalog and the NED database reveals the
presence of an additional 26 galaxies with concordant redshifts
(within $3\sigma_v$ of the group mean) and within $\sim 1.5$~Mpc from
the X-ray peak, some of which may represent group members infalling
for the first time. For these 60 system members, we derive a mean
redshift of $z=0.0607\pm 0.0002$ and a velocity dispersion $\sigma_v =
503^{+56}_{-71}$~km~s$^{-1}$ \citep{bai10}. Table~\ref{tab,x_opt}
summarizes relevant X-ray and optical properties of the system derived
here and in \citet{bai10}. Remarkably, while our {\em XMM} data show
symmetrically distributed hot gas out to several hundred kpc (see
Section~\ref{sec,results}), consistent with a virialized system, the
core of the system is dominated by {\em two} optically luminous
early-type galaxies, as shown in Figure~\ref{fig,magellan}.  This
central galaxy pair was previously noted by \citet{arp87} in their
catalog of southern peculiar galaxies. In the following, the optically
brighter northern galaxy in the pair will be referred to as Galaxy~A
and its southern counterpart as Galaxy~B.

\begin{table}
\caption{Global properties of MZ\,10451\label{tab,x_opt}}
\begin{center}
\begin{tabular}{lcc}
  \tableline \hline
  & Value & Source \\ \hline
 $\langle z \rangle$              & $0.0607\pm 0.0002$ & Magellan/2dFGRS  \\
 $N_{\rm gal}$                    & 60     & Magellan/2dFGRS  \\
 $\sigma_v$         & $503^{+56}_{-71}$~km~s$^{-1}$  & Magellan/2dFGRS \\ 
 $r_{\rm X}$              & 430~kpc & {\em XMM} \\
 $L_{\rm X}$ & $2.2\pm 0.3\times 10^{42}$~erg~s$^{-1}$ & {\em XMM} \\
 $\langle T_{\rm X}\rangle$ & $0.91^{+0.06}_{-0.08}$~keV &  {\em XMM}  \\
 $r_{500}$ & $329\pm 12$~kpc & {\em XMM}/{\em Chandra} \\
 $r_{200}$ & $491^{+17}_{-18}$~kpc & {\em XMM}/{\em Chandra}\\
 $r_{\rm vir}$ & $661^{+21}_{-23}$~kpc &  {\em XMM}/{\em Chandra} \\
 $M_{500}$     & $1.3\pm 0.1\times 10^{13}$~M$_\odot$  & {\em XMM}/{\em Chandra}  \\
 $M_{200}$     & $1.6\pm 0.1\times 10^{13}$~M$_\odot$  & {\em XMM}/{\em Chandra}  \\
 $M_{\rm vir}$ & $1.9\pm 0.2\times 10^{13}$~M$_\odot$  & {\em XMM}/{\em Chandra}  \\
  \tableline
\end{tabular}
\tablecomments{Optical properties taken from \citet{bai10}. $r_{\rm
  X}$ is the radius of detectable X-ray emission, $\langle T_{\rm
  X}\rangle$ the emission-weighted mean ICM temperature within $r_{\rm
  X}$, and $r_{\rm vir}$ is here identified with $r_{100}$ in the
  adopted cosmology.}
\end{center}
\end{table}

\begin{figure}
\begin{center}
\epsscale{1.16} 
\plotone{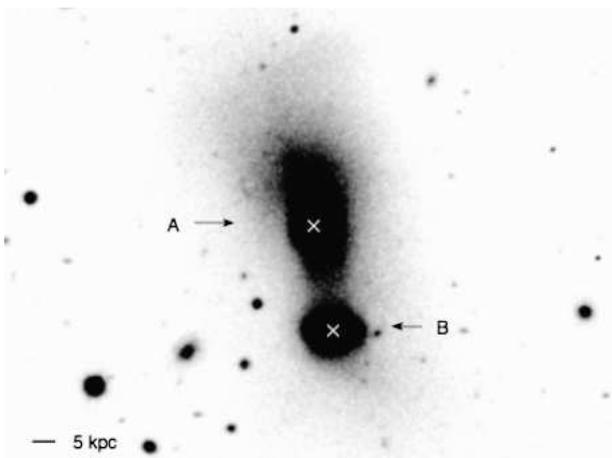}
\end{center}
\figcaption{Magellan $B$-band image of the core of MZ\,10451, taken at
   $\sim 0.6''$ seeing, with Galaxies A and B labeled and their
   optical centers marked by crosses. Display levels have been chosen
   to emphasize the faint stellar components.
\label{fig,magellan}}
\end{figure}

There is considerable evidence to suggest an ongoing merger between
the two galaxies. Firstly, their optical centers are separated by only
$20''$ ($\sim 25$~kpc) in projection, and their radial velocity
offsets relative to the group mean are only 34~km~s$^{-1}$ (Galaxy A)
and 2~km~s$^{-1}$ (B). This is well within the typical uncertainty in
our IMACS redshift measurements of $\sim 50$~km~s$^{-1}$
\citep{rasm06}, demonstrating that, along the line of sight, both
galaxies are consistent with being at rest relative to the group
center. Secondly, both galaxies display morphological features
consistent with an interaction. Most prominent is an extensive stellar
plume that extends $\sim 50$~kpc northwards from the optical nucleus
of Galaxy~A. An extended stellar component is also visible around
Galaxy~B, as is a bright stellar bridge between the two objects.

\section{OBSERVATIONS AND ANALYSIS}\label{sec,obs}

MZ\,10451 was observed by {\em Chandra} in Very Faint telemetry mode
for a total exposure of $\sim 80$~ks, with the ACIS-S3 CCD as
aimpoint. Observations were split into three separate pointings, which
were analyzed using {\sc ciao} v4.1. To apply the newest calibration
data and enable the suppression of background events afforded by the
telemetry mode, new level~2 event files were generated. CTI
corrections, time-dependent gain corrections, and standard grade
filtering was applied. Lightcurves extracted for the S3 CCD in the
2.5--7~keV band revealed no evidence for background flares in any of
the pointings, leaving a total effective exposure time of 78.56~ks.
While the focus here is on the properties of the central group regions
of MZ\,10451 as seen by {\em Chandra}, we facilitate the analysis by
including our {\em XMM} data where relevant. These data were prepared
following the prescription outlined in \citet{rasm06}.

The {\em XMM} data indicate detectable ICM emission out to at least
$r\approx 6'$ from the X-ray peak (Section~\ref{sec,results}),
implying that any methods involving local background estimation should
be avoided for the analysis of diffuse emission in the {\em Chandra}
S3 data. Employing the ``blank-sky'' background data in the {\em
Chandra} calibration database for this purpose is also problematic,
because the particle background component is now much more pronounced
than at the time of these blank-sky observations (see, e.g., the
discussion in \citealt{sun09}). We have verified this by extracting
blanksky-subtracted source spectra in regions that are not
source-dominated, revealing prominent excess emission even at $E<
2$~keV, presumably from unsubtracted particles.  Instead, we employed
the 275-ks of `period~E' background data taken with the ACIS array in
the stowed position, in which only non-X-ray events are
recorded. These data can thus be used to estimate the particle
contribution to the total background.  The latest observation to be
included in these data was performed just three months prior to our
MZ\,10451 observation and so should not display serious discrepancies
with respect to the actual particle level in our data. Indeed, in the
9.5--12~keV band, the total flux ratio between source and stowed data
exceeds unity by an acceptable 7--9\% among our three pointings
(compared to almost 60\% for to blanksky data). The stowed data are
scaled up by these small factors in our analysis.

In all spectral analysis, we subtracted the particle contribution
using the stowed data, and modeled the remaining background
components. The latter include the extragalactic X-ray background,
whose spectrum can be well described by a power-law of photon index
$\Gamma \approx 1.5$, and diffuse Galactic emission which is strongly
spatially varying but can be described by a low-temperature thermal
plasma model. Our {\em XMM} data were used to constrain the properties
of these two components at the position of MZ\,10451 as follows.
Spectra were extracted for each EPIC camera in bins of $\ge 50$~net
counts within a $7.5'$--10$'$~annulus of the X-ray peak. Energies
$>1.4$~keV were excluded, to suppress the particle-induced
contribution including the instrumental fluorescence lines at
$E=1.4$--1.6~keV.  The results were jointly fitted with a model
consisting of a power-law subject to Galactic absorption plus an
unabsorbed solar-abundance thermal plasma model. Free parameters were
thus the power-law index $\Gamma$, plasma temperature $T_{\rm Gal}$,
and the normalization of the two components. The resulting fit, with
$\chi^2_\nu = 1.36$ for 72 degrees of freedom, yielded uncertainties
on the best-fit values of $\Gamma=1.54$ and $T_{\rm Gal}=0.12$~keV of
only $\sim 5$\%.  The pn spectrum and best-fit model are presented in
Figure~\ref{fig,pnbkg}. We include this background model in all fits
to the {\em Chandra} data. Leaving the relative normalization of the
thermal and power-law components free to vary did not impact
significantly on our results, so this ratio was fixed to that derived
from the {\em XMM} fit, with only the overall background normalization
remaining as a free parameter.

\begin{figure}
\begin{center}
\epsscale{1.05} 
\plotone{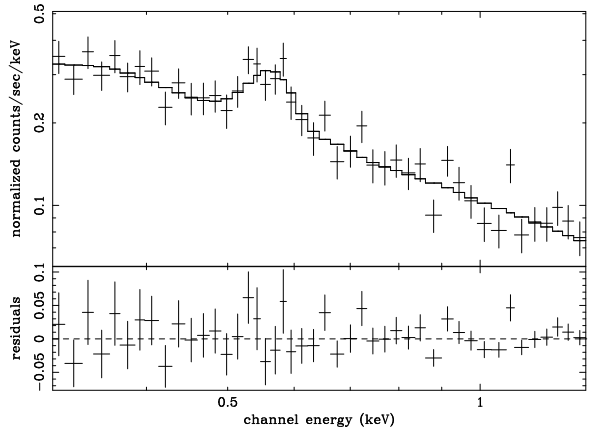}
\end{center}
\figcaption{{\em XMM} pn soft X-ray background spectrum extracted
    within a 7.5$'$--10$'$~annulus centered on the X-ray peak, along
    with the best-fit model. Bottom panel shows fit residuals.
\label{fig,pnbkg}}
\end{figure}

Source and stowed-background spectra and associated response products
were extracted separately for each of the three {\em Chandra}
pointings, with spectra accumulated in bins of $\geq 20$~net~counts
and jointly fitted in {\sc xspec} v11.3. Whenever relevant, a similar
approach was adopted for source and blanksky-background spectra
extracted for each {\em XMM} camera.

In addition to the X-ray data, we also acquired optical long-slit
spectroscopy of each of the merging galaxies, using the IMACS camera
at the Baade/Magellan telescope in November 2009. The f/2 camera mode
was used with the 300~lines~mm$^{-1}$ grism, giving a nominal
wavelength range of 3900--10000~{\AA} and a dispersion of
1.34~{\AA}~pixel$^{-1}$. Each galaxy was observed for 30~min using a
0.9-arcsec wide slit (corresponding to a physical scale of $\sim$
1~kpc at the group redshift).  General details of the flat-fielding,
bias- and sky-subtraction, and wavelength calibration can be found in
\cite{rasm06}.

\section{RESULTS}\label{sec,results}

\subsection{Global and Radial ICM Properties}

The large-scale morphology of the ICM in MZ\,10451 is most easily
discerned from our {\em XMM} data. Based on a particle-subtracted and
exposure-corrected 0.3--2~keV EPIC mosaic image,
Figure~\ref{fig,xmmmosaic}(a) shows adaptively smoothed {\em XMM}
contours of the central $8'\times 8'$~arcmin ($\sim 0.5\times
0.5$~Mpc$^2$) region. The surface brightness distribution clearly
exhibits a fairly symmetric morphology, consistent with the
expectation for a virialized system. On the largest scales probed by
{\em XMM}, there is an indication of the ICM distribution being
slightly elongated in the north--south direction, roughly coinciding
with the axis joining the two central galaxies.

\begin{figure}
\begin{center}
\epsscale{0.90} 
\plotone{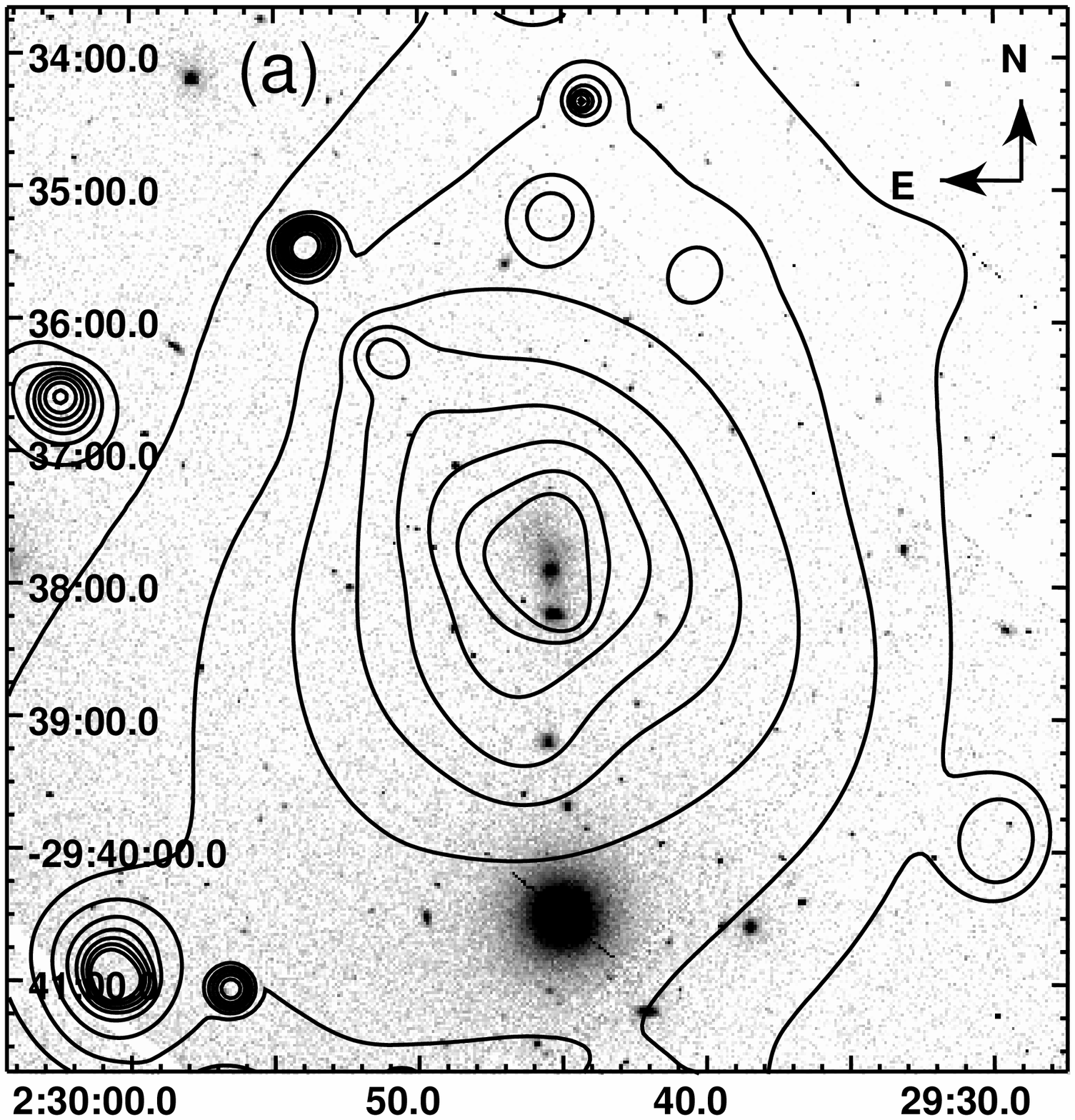}\\
\epsscale{1.015} 
\mbox{\hspace{-11mm}
\plotone{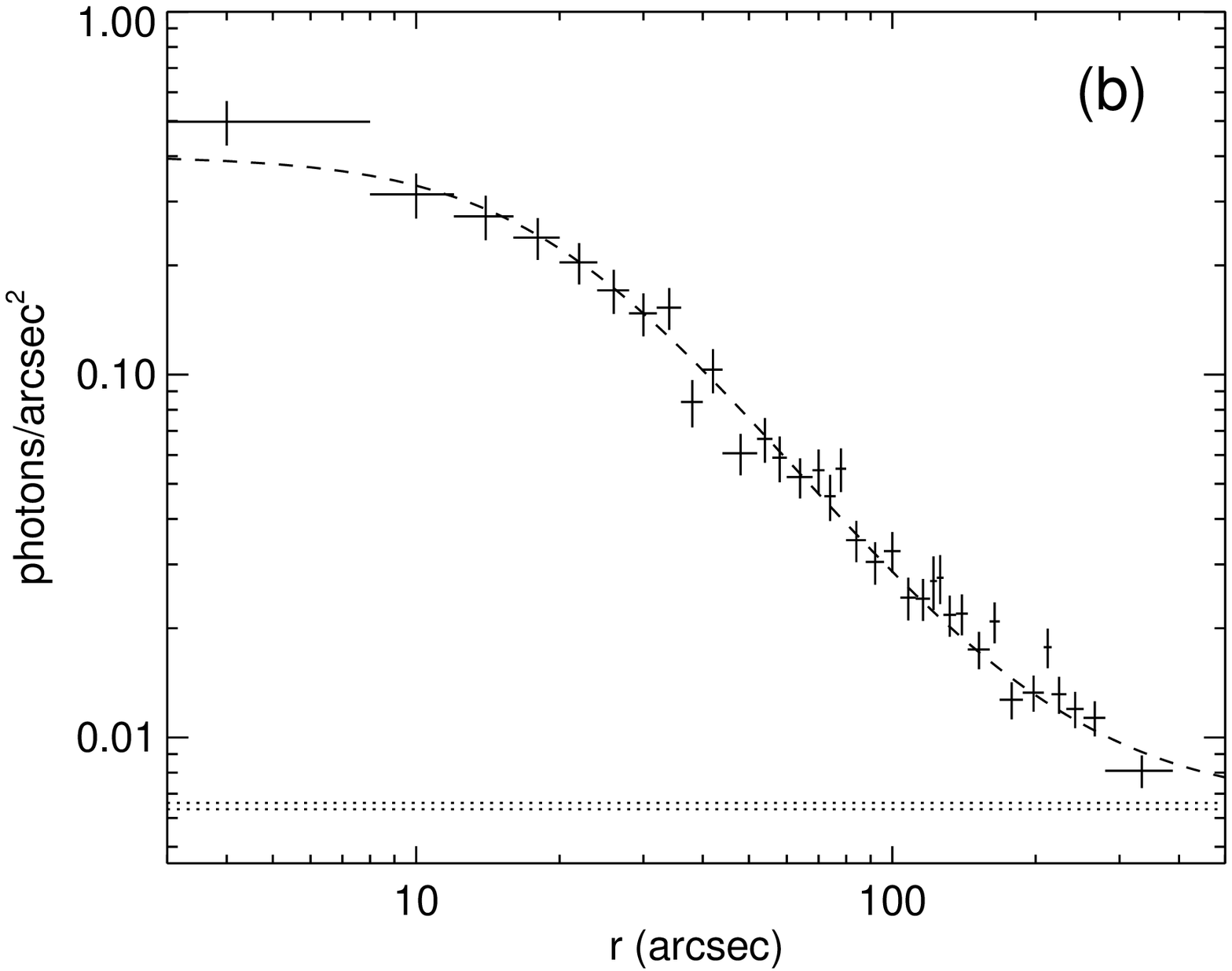}}
\end{center}
\figcaption{(a) {\em XMM} 0.3--2~keV surface brightness contours
  overlayed on our Magellan $R$--band image of MZ\,10451. (b)
  Corresponding radial surface brightness profile. Dashed line shows
  the best-fit $\beta$--model, with $\beta=0.46\pm 0.01$ and $r_c =
  20.4\pm 1.7''$, while dotted horizontal lines outline the 1-$\sigma$
  uncertainties on the local background level.
\label{fig,xmmmosaic}}
\end{figure}

Based on the same (unsmoothed) data, Figure~\ref{fig,xmmmosaic}(b)
shows a radial surface brightness profile of the diffuse emission,
extracted from the X-ray peak in bins containing a signal-to-noise
ratio of S/N~$\ge 10$. Uncertainties in this plot contain a 10\%
systematic error associated with the subtraction of the {\em XMM}
particle background, added in quadrature to all Poisson
uncertainties. Emission is robustly detected out to at least $r_{\rm
X}\approx 390$~arcsec ($\approx 430$~kpc). A simple $\beta$--model
with $\beta =0.46$ provides a reasonable description of the profile
across this radial range, with the exception of the very core which
shows a slight excess of emission above the best-fit model. A thermal
plasma model (APEC) fit to the global {\em XMM} spectrum within the
full region covered in Figure~\ref{fig,xmmmosaic}(b) provides an
acceptable fit, $\chi^2_\nu=1.06$ for 111 degrees of freedom, and
would suggest $T=0.91^{+0.06}_{-0.08}$~keV and
$Z=0.09^{+0.03}_{-0.02}$~Z$_\odot$.  However, we note that this
abundance measurement is plausibly biased low due to the likely
presence of temperature variations across this large radial range ("Fe
bias"; \citealt{buot00}). The best-fit model would imply a 0.3--2~keV
luminosity of $L_{\rm X} = 2.2 \pm 0.3 \times 10^{42}$~erg~s$^{-1}$,
with the bolometric value a factor of 1.8 higher.

Subtracting the particle background as estimated from stowed data and
excluding identified point sources, a total of $\sim$\,4,900 net
counts (0.3--2~keV) from diffuse X-ray emission is detected on the
{\em Chandra} S3~CCD, roughly a factor of four more than in the
shallower {\em XMM} pn data of the same region.
Figure~\ref{fig,cxomosaic} shows a particle-subtracted,
exposure-corrected, and adaptively smoothed {\em Chandra} mosaic image
of the central regions. Afforded by the superior spatial resolution of
{\em Chandra}, these data clearly demonstrate that both central
galaxies are offset from the peak of the diffuse X-ray emission at
($\alpha$, $\delta$) = (02$^{\rm h}29^{\rm m}45\fs4$, $-29\degr
37\arcmin 43$), suggesting that neither galaxy is completely at rest
at the center of the group gravitational potential. The X-ray
morphology of the group core is quite regular, presenting little
evidence that the ongoing galaxy interaction has yet had any profound
impact on the central ICM density distribution.  Also note that an
emission component at the optical position of Galaxy~B is clearly
visible.

\begin{figure}
\begin{center}
\epsscale{1.16} 
\plotone{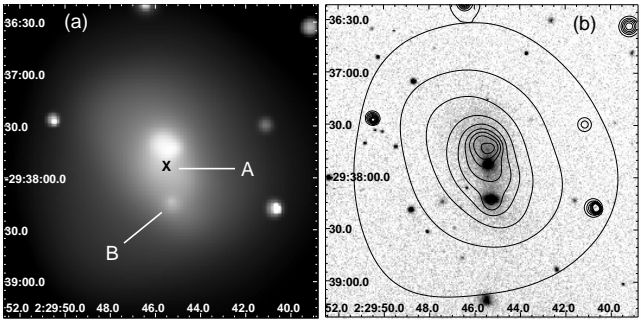}
\end{center}
\figcaption{(a) 0.3--2~keV smoothed {\em Chandra} image of the central
   $3'\times 3'$ around the X-ray peak, with the optical centers of
   the two merging galaxies marked. (b) Contours of the left image
   overlayed on Magellan $R$--band image.
\label{fig,cxomosaic}}
\end{figure}

To test for radial variations in ICM temperature and abundance and
search for ICM signatures of the ongoing merger, {\em Chandra} spectra
were extracted in concentric annuli centered on the X-ray peak, each
containing $\sim$1,000~counts from diffuse X-ray emission. These were
fitted with an APEC model added to the X-ray background model
described in Section~\ref{sec,obs}. Deprojection was not attempted, as
our {\em XMM} data show that the regions beyond the {\em Chandra} S3
CCD are not free of group emission. The resulting radial profiles of
$T$ and $Z$ are shown in Figure~\ref{fig,profiles}. Note that the
X-ray peak is located $\sim 1.4'$ east of the S3 CCD center, enabling
coverage towards the western CCD corners to $r\approx 6'$ in
Figure~\ref{fig,profiles}, although the angular coverage in the
outermost radial bin is considerably less than 360~degrees.

\begin{figure}
\begin{center}
\epsscale{1.05} 
\plotone{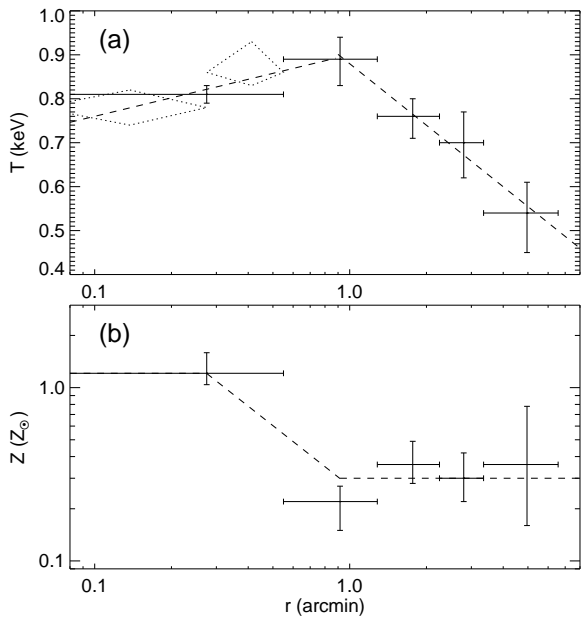}  
\end{center}
\figcaption{{\em Chandra} profiles of (a) ICM temperature and (b)
   abundance.  Dotted diamonds in (a) show the fit results if further
   subdividing the innermost radial bin while keeping $Z$ fixed at the
   value in the parent bin. Dashed lines show our parameterizations of
   the profiles.
\label{fig,profiles}}
\end{figure}

The temperature profile reveals the presence of slightly cooler gas in
the group core, showing a peak in the radial bin that extends from
$\sim 0.6'$--1.3$'$ ($\sim 40$--90~kpc) from the core. The central
temperature drop may not be quite as pronounced as in typical
cool-core groups, with the innermost sub-bin in
Figure~\ref{fig,profiles} nominally being only $\sim 15$\% cooler than
the peak temperature (the drop is significant at the $2.0\sigma$
level). However, resolution effects could have an impact on this,
given the higher redshift of this system compared to those of typical
well-studied X-ray groups (see, e.g., \citealt{mulc03}). An orthogonal
regression fit in $T$--log\,$r$ space to the outermost four bins in
Figure~\ref{fig,profiles}(a) yields $\alpha = {\rm d}T/{\rm d log\,}r
= -0.47\pm 0.02$ for the slope outside the core, consistent with
results for typical cool-core groups \citep{rasm07}.  We can therefore
approximate the temperature profile as a piecewise log--linear
function,
\begin{equation}
  \mbox{$T(r) =
  \cases{+0.14\mbox{\,log\,}(r/\mbox{kpc}) + 0.65 , \,\,\,\, r\leq 0.9'  \cr 
         -0.47\mbox{\,log\,}(r/\mbox{kpc}) + 1.73 , \,\,\,\, r> 0.9'}$}. 
\label{eq,tprof} 
\end{equation}
Similarly, the abundance profile can be approximated as constant at
0.3~solar outside the core, rising linearly in log\,$r$--log\,$Z$
space to its central value. These parametrizations are illustrated by
the dashed lines in Figure~\ref{fig,profiles}.

While the temperature profile is thus typical of cool-core groups, the
abundance profile is slightly unusual. Although showing the central
peak of roughly solar abundance typical of such systems, the profile
does not exhibit the steady radial decline seen in such groups but is
instead consistent with being largely flat outside the core. Note also
that the derived {\em Chandra} abundances are everywhere larger than
suggested by the fit to the global {\em XMM} spectrum within this
region, supporting the suspicion that the {\em XMM} estimate could be
affected by the Fe bias arising from the presence of significant
temperature variations.  We also checked the Si/Fe ratio for the
region encompassing the innermost two bins in
Figure~\ref{fig,profiles}(b), using a VAPEC model fit.  Although
$Z_{\rm Si}$ is poorly constrained, the ratio is found to be
$<0.9$~Z$_{{\rm Si},\odot}/$Z$_{{\rm Fe},\odot}$ at $1\sigma$
significance, indicating an important contribution to central
enrichment from type~Ia supernovae and hence from an old stellar
population.

Combining the {\em XMM} surface brightness profile, which is well
constrained to large radii, with the parametrized {\em Chandra}
temperature and abundance profiles (slightly smoothed to ensure a
continuous behavior), profiles of deprojected gas density $n_e$,
entropy $S=T/n_e^{2/3}$, cooling time, and total mass were
determined. The latter was obtained via the assumption of hydrostatic
equilibrium, as supported by the regular ICM morphology on large
scales. The results are plotted in Figure~\ref{fig,mass}.
Uncertainties on each profile were obtained from 1,000 Monte Carlo
realizations, for each of which all relevant parameters ($T(r)$,
$Z(r)$, $\beta$, $r_c$, and surface brightness normalization) were
drawn from a Gaussian distribution centered at the measured or
parametrized value and with 1-$\sigma$ width equal to the typical
1-$\sigma$ uncertainty on the relevant parameter. At each interpolated
radius, we recorded the resultant outlier-resistant biweight mean and
standard deviation. The latter represent the 1-$\sigma$ error bounds
shown in Figure~\ref{fig,mass}. Errors on overdensity radii were
evaluated from the mass profile using an analogous procedure.

\begin{figure}
\begin{center}
\epsscale{1.17} 
\plotone{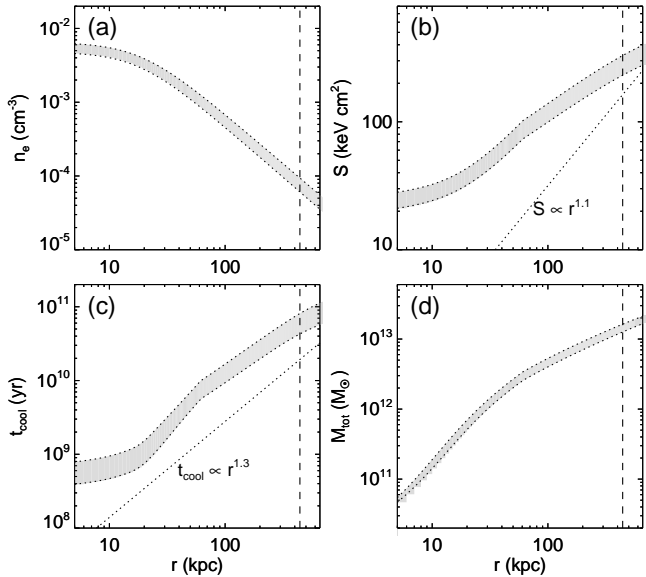}
\end{center}
\figcaption{Radial profiles of (a) ICM electron density, (b) entropy,
  (c) cooling time, and (d) cumulative total mass, within the
  estimated virial radius. Shaded regions represent $1\sigma$
  uncertainties. Dashed vertical lines outline the X-ray detection
  radius in the {\em XMM} data.  Dotted line in (b) shows the
  behaviour expected from gravitational heating only (arbitrarily
  normalized), while the line in (c) represents the empirical cluster
  relation from \citet{sand06}, also arbitrarily normalized.
\label{fig,mass}}
\end{figure}

The entropy profile is fairly typical of cool-core groups, showing a
continuous decline towards the center, and a logarithmic slope at
large radii which is flatter than that expected from gravitational
processes only \citep{john09, sun09}. The cooling time profile shows a
central value of $0.6\pm 0.2$~Gyr and a slope at large $r$ in good
agreement with results for hotter systems \citep{sand06}. The derived
mass profile implies a total mass of $M = 1.4\pm 0.1 \times
10^{13}$~M$_\odot$ within the radius of X-ray detection ($r \approx
6.5' \approx 430$~kpc), which is itself bracketed by the resulting
values of $r_{500} = 329\pm 12$~kpc and $r_{200}=491^{+17}_{-18}$~kpc.
The estimated virial radius in the adopted cosmology is $r_{100} =
661^{+21}_{-23}$~kpc, enclosing an extrapolated mass of $M_{\rm vir} =
1.9 \pm 0.2 \times 10^{13}$~M$_\odot$. Table~\ref{tab,x_opt} includes
relevant quantities obtained from the mass profile.

\subsection{2-D ICM Maps and Spectroscopy}

Although MZ\,10451 is intrinsically fairly X-ray luminous, its
considerable distance ($\sim 260$~Mpc) still renders it too faint to
allow spectral mapping at fine spatial detail with the present {\em
Chandra} data. Hence, to further investigate the ICM properties in the
group core and to identify regions of particular interest for spectral
analysis, we first generated a hardness ratio map of the group core.
For this, energy bands of 0.5--1 and 1--2~keV were chosen, in order to
provide comparable number of counts in the `soft' and `hard' bands. We
avoid the lowest energies, to reduce contamination from any patchy,
soft Galactic emission. Mosaic images in each band were
particle-subtracted and corrected for exposure variations, and results
were smoothed using scales resulting from adaptively smoothing the
corresponding full-band 0.5--2~keV image. The hard-band image was then
divided by the soft-band one. For the temperatures and abundances seen
in Figure~\ref{fig,profiles}, the hardness ratio map $H$ in these
energy bands provides a useful proxy for spatial variations in ICM
temperature $T^\prime$ along the line of sight \citep{fino06}.
Utilizing this, 2-D maps of projected pseudo-entropy $S^\prime$ and
pressure $P^\prime$ were also constructed, based on $S^\prime \sim H/
I_e^{1/3}$ and $P^\prime \sim HI_e^{1/2}$, where $I_e = \int n^2
\,\mbox{d}l$ is the ICM emission measure. To generate these maps, a
0.5--2~keV image, smoothed on the same scales as $H$, was taken as a
proxy for $I_e$ (modulo metallicity variations), since at fixed
metallicity $Z \sim 0.3$~Z$_\odot$ (as seen outside the very core),
the emissivity of a plasma varies by $<5$\% for the range of
temperatures in Figure~\ref{fig,profiles} \citep{suth93}.
Nevertheless, caution should be exercised when interpreting these
maps, due to the degeneracy with metallicity and the presence of
statistical fluctuations.

The results are displayed in Figure~\ref{fig,maps}, which shows the
central $2.5'\times 2.5'$ region around the X-ray peak, corresponding
to the region covered by the two innermost bins in
Figure~\ref{fig,profiles}(b). The hardness ratio map in
Figure~\ref{fig,maps}(a) generally suggests fairly uniform hardness
ratios on these spatial scales, but with some notable exceptions: The
central regions display relatively soft emission, while a region of
slightly harder emission is seen south of the group core. If
interpreting these features as due to variations in ICM temperature,
the results support the presence of slightly cooler gas in the group
core on scales of 20--30~kpc, as already hinted at by
Figure~\ref{fig,profiles}. Several of the detected point sources in
this region also show emission which is considerably harder than that
of the ICM, but, interestingly, this is less obviously so for the two
central galaxies. The map also suggests the presence of a curved
region of slightly hotter material immediately south of Galaxy~B,
referred to in Figure~\ref{fig,maps}(a) and in the following as the
southern ``hotspot''.

\begin{figure*}
\begin{center}
\epsscale{1.0}
\plotone{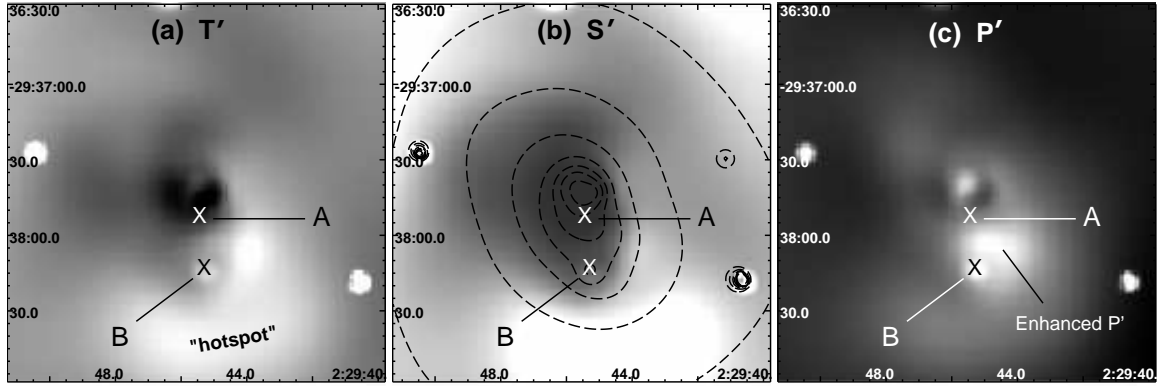}
\end{center}
\figcaption{(a) {\em Chandra} (1--2~keV)/(0.5--1~keV) hardness ratio
    map of the central $2.5'\times 2.5'$ region, with the optical
    centers of the galaxy pair marked. Values range from 0.4 (black)
    to 1.0 (white); for a $Z\approx 0.3$~Z$_\odot$ plasma at $z=0.06$
    subject to Galactic absorption, this corresponds to $T\approx 0.8$
    and 1.2~keV, respectively. (b) Projected entropy, with surface
    brightness contours from Figure~\ref{fig,cxomosaic} overlayed for
    comparison.  (c) Projected pressure. Normalization in (b) and (c)
    is arbitrary, with dark colors indicating low values.
\label{fig,maps}}
\end{figure*}

The entropy map in Figure~\ref{fig,maps}(b) suggests the presence of a
central region of relatively low-entropy gas, centered close to the
X-ray peak. The entropy distribution generally appears fairly
symmetric and reveals no prominent features around the two central
galaxies. In particular, there is no clear indication of ongoing ISM
stripping from either galaxy, e.g., in the form of irregularly
distributed low-entropy material in their immediate vicinity. Note
that the southern hotspot visible in the hardness ratio map also shows
up in this diagram as a region of slightly enhanced ICM entropy.

The pressure map, Figure~\ref{fig,maps}(c), indicates that ICM thermal
pressure is generally declining outwards from the X-ray peak, as
expected for hot gas in approximate hydrostatic equilibrium. However,
the map also presents several potentially interesting features. One of
these is a region of seemingly enhanced ICM pressure {\em between} the
two galaxies, as marked in the Figure. No corresponding surface
brightness enhancement is clearly seen (which could suggest a local
enhancement in $Z$ at fixed projected density), nor in entropy or
hardness ratio. We also note that the X-ray peak seems to be
surrounded by two small low-pressure regions, with no obvious
counterparts in the surface brightness distribution (cf.\
Figure~\ref{fig,cxomosaic}), but the available statistics do not allow
a detailed investigation of their robustness or possible origin.

We can, however, test the robustness of the hotspot seen south of
Galaxy~B. For this, a spectrum was extracted in a rectangular region
covering this feature and fitted with a thermal plasma model.
Figure~\ref{fig,specregions} outlines the spectral extraction regions
discussed in the following. Within the region labelled ``H'' in the
Figure, the derived temperature, $T = 0.98 \pm 0.06$~keV, is indeed
higher than that obtained for a comparable control region at similar
distance north of the X-ray peak (region ``C'' in
Figure~\ref{fig,specregions}), $T = 0.85^{+0.06}_{-0.05}$~keV.
Formally, these results differ at 1.5-$\sigma$ significance, with the
southern region being $\sim 15$\% hotter. We note that although the
position of the temperature peak in Figure~\ref{fig,profiles} is
consistent with that of this region, results for the radial profile do
not change significantly if excluding this region from the second
radial bin in the profile; the presence of a peak in the temperature
profile at this radius is therefore not simply due to the hotspot. We
also note that the derived metallicity within this region,
$Z=0.27^{+0.10}_{-0.08}$~Z$_\odot$, is fully consistent with the
radial average at this distance from the X-ray peak,
$Z=0.22^{+0.05}_{-0.07}$~Z$_\odot$. This implies that the presence of
this feature is not simply due to strong local variations in ICM
abundance.

\begin{figure}
\begin{center}
\epsscale{1.0} 
\plotone{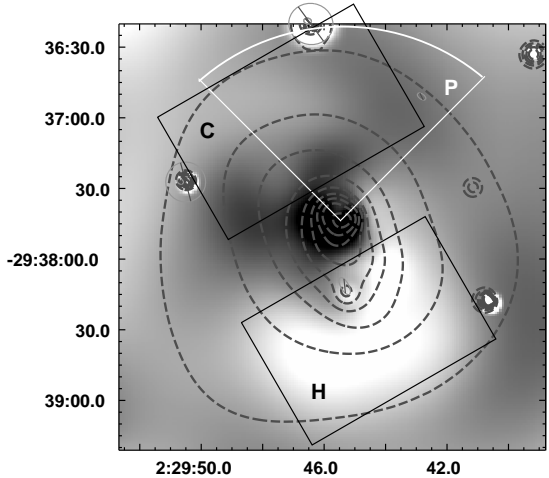}
\end{center}
\figcaption{Spectral extraction regions employed for the southern
   hotspot (``H''), the corresponding control region (``C''), and the
   stellar plume of Galaxy~A (``P''). For reasons of clarity, control
   regions for the latter are not shown (see text). Regions are
   overlayed on a hardness ratio map of the central $3'\times 3'$
   region, with contours from Figure~\ref{fig,cxomosaic}.
\label{fig,specregions}}
\end{figure}

Another interesting question is how the ICM metallicity in the region
covered by the stellar plume of Galaxy~A compares to that of its
immediate surroundings. This plume might represent tidally stripped
material that could evolve into an intracluster light component. These
stars would be able to chemically pollute the surrounding ICM very
efficiently, as their ejecta would mix directly with the ICM without
having to overcome the confining gravitational potential and gas
pressure of their former host galaxy. To test this possibility, a
spectrum was extracted in a wedge extending $1.5'$ northwards from the
X-ray peak (region ``P'' in Figure~\ref{fig,specregions}), with the
minimum size of the region dictated by the need to get at least 500
counts as required for a rough measurement. Corresponding wedges
extending to the east and west of the peak were employed as control
regions. All regions had their apex centered on the X-ray peak rather
than on Galaxy~A itself, such that the metal-rich group core (cf.\
Figure~\ref{fig,profiles}) would be equally represented in all wedges
and thus not bias the result for the northern wedge. For the N wedge,
we find $Z=0.65^{+0.28}_{-0.14}$~Z$_\odot$, compared to $0.19\pm
0.05$~Z$_\odot$ for the combined E and W wedges. Hence, gas coinciding
with the stellar plume does show significantly higher abundance than
other regions at similar distance from the group core. This supports a
scenario in which the central rise in the radial abundance profile has
a significant contribution from highly enriched gas coinciding with
this plume.

\subsection{The Central Galaxy Pair}\label{sec,pair}

Turning now to the properties of the interacting galaxies, an
important question is whether the galaxies have retained any hot gas,
and whether the ongoing interaction has triggered significant
starburst or active galactic nucleus (AGN) activity in
either. Galaxy~B is particularly interesting in this regard, because
our {\em Spitzer} data reveal $24\,\mu$m emission from this galaxy
\citep{bai10}. To explore these issues, we present Magellan optical
spectra of both galaxies in Figure~\ref{fig,opt_spec} (obtained with a
slit width covering the central $\sim 1$~kpc of each galaxy), along
with our near-UV and $24\,\mu$m data of the pair in
Figure~\ref{fig,galpair}. The optical spectra appear remarkably
similar, and neither shows clear evidence for bright emission lines
typically associated with star formation or nuclear activity. This is
in stark contrast to results showing that practically all (fully
formed) BCGs close to the X-ray center in cool-core clusters display
optical line emission \citep{edwa07}. Also note that the nucleus of
Galaxy~B is actually the brighter at most optical wavelengths, whereas
that of Galaxy~A dominates towards the near-UV as also confirmed by
our {\em GALEX} data.

\begin{figure}
\begin{center}
\epsscale{1.1} 
\plotone{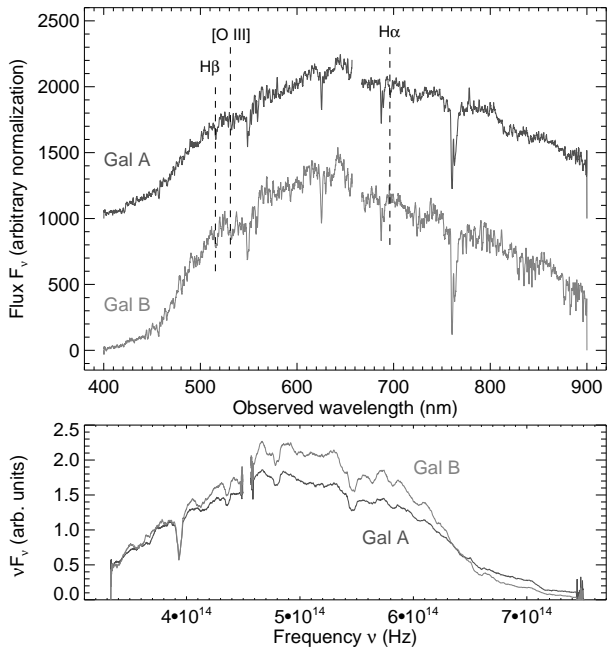}
\end{center}
\figcaption{Magellan long-slit optical spectra of Galaxy~A and B,
  offset for clarity. Wavelengths are in the observer frame, with the
  location of typical bright emission lines labeled.  The feature
  around 759~nm is due to atmospheric O$_2$ A-band absorption and the
  gap at 657--667~nm to a chip gap in the IMACS camera. Bottom panel
  shows a heavily smoothed $\nu F_\nu$ representation (with no
  offset).
\label{fig,opt_spec}}
\end{figure}

\begin{figure}
\begin{center}
\epsscale{1.16} 
\plotone{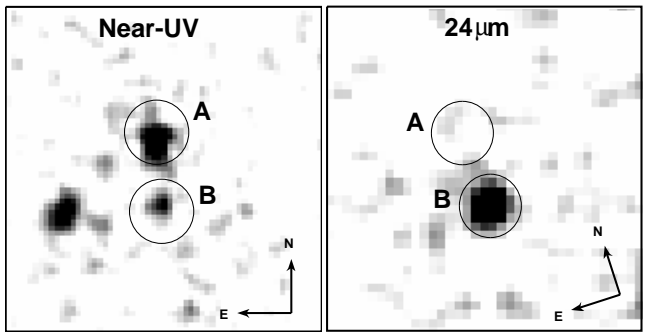}
\end{center}
\figcaption{{\em GALEX} NUV and {\em Spitzer} $24\,\mu$m images of the
  central $1.5'\times 1.5'$ around the galaxy pair, with circles
  marking their optical centers. Images have been smoothed with
  Gaussians of $\sigma = 3\arcsec$ (NUV) and $5\arcsec$
  ($24\,\mu$m). Both galaxies are detected by {\em GALEX}, whereas
  only Galaxy~B is also confirmed as a $24\,\mu$m source.
\label{fig,galpair}}
\end{figure}

Given the lack of bright optical emission lines in Galaxy~B, its
$24\,\mu$m emission may be attributed to either obscured AGN activity
or dusty star formation. In contrast, the optically brighter Galaxy~A
is not detected above $3\sigma$ significance at
$24\,\mu$m. Nevertheless, extended emission from both galaxies is
detected in the {\em GALEX} NUV band at $>5\sigma$ significance. This
may suggest the presence of some recent or ongoing low--level star
formation, although the NUV--$R$ colors of 5.5 (Galaxy~A) and 6.1 (B)
would render both galaxies "quiescent" by some definitions (e.g.\
\citealt{cort08}). In addition, based on $B_J$ magnitudes from the 2dF
survey, both galaxies have comparable optical colors of
$B_J$--$R=2.0\pm 0.1$ (A) and $2.1\pm 0.1$ (B), consistent with
typical passive early-types.  Neither galaxy is detected in the
far-UV, implying FUV--NUV\,$\gtrsim 1$ in either case, nor at
$70\,\mu$m in our {\em Spitzer} data. They are also not detected at
1.4-GHz in NVSS radio data, which are 99\% complete down to
$S_{21\rm{cm}} = 3.4$~mJy \citep{cond98}, implying 1.4-GHz
luminosities below $2.8\times 10^{22}$~W~Hz$^{-1}$.  Finally, to
estimate their stellar masses, we note that 2MASS isophotal photometry
is available for Galaxy~B only, implying $L_K=9.6\times
10^{10}$~L$_\odot$ for this galaxy. If assuming a $K$-band stellar
mass-to-light ratio $M_\ast/L_K \approx 1$~M$_\odot/$L$_\odot$, then
$M_\ast/L_R \approx L_K/L_R \approx 5.3$~M$_\odot/$L$_\odot$ for
Galaxy~B. Assuming a similar relationship for Galaxy~A (see also
\citealt{liu09}) yields the approximate stellar masses listed in
Table~\ref{tab,gals} which summarizes some relevant properties of the
two galaxies. Note that the stellar mass estimate for Galaxy~A
includes the diffuse stellar plume.

\begin{table}
\caption{Salient parameters for the central galaxy pair
\label{tab,gals}}
\begin{center}
\begin{tabular}{lcc}
 \tableline \hline
 & Galaxy~A & Galaxy~B \\ \hline
RA (J2000)                        & 02$^{\rm h}29^{\rm m}45\fs34$    &  02$^{\rm h}29^{\rm m}45\fs23$  \\
Dec (J2000)                       & $-29\degr 37\arcmin 51\farcs 2$  & $-29\degr 38\arcmin 13\farcs 4$ \\
$L_R$ (L$_\odot$)                 & $3.6\times 10^{10}$ & $1.8\times 10^{10}$\\
$L_{\rm IR}$  (L$_\odot$)         & $<8\times 10^8$     & $2.0\times 10^9$ \\
$m_{\rm NUV}$                     & 20.62 & 21.97 \\
$L_{\rm X}(<3'')$ (erg~s$^{-1}$)  & $2.0\pm 0.4\times 10^{40}$ &
                                    $2.0\pm 0.4\times 10^{40}$ \\
$L_{\rm 1.4\,GHz}$ (W Hz$^{-1}$)  & $<2.8\times 10^{22}$&$<2.8\times 10^{22}$\\
$M_\ast$ (M$_\odot$) & $1.9\times 10^{11}$ & $1.0\times 10^{11}$ \\
  \tableline
\end{tabular}
\tablecomments{Coordinates indicate the optical galaxy
  centers. Infrared luminosities are based on our {\em Spitzer}
  $24\,\mu$m data, NUV magnitudes are from our {\em GALEX} data
  (uncorrected for dust attenuation), and X-ray luminosities within
  $r=3''$ of the optical center are given in the 0.5--8~keV
  band. Radio luminosities are upper limits from NVSS data, and
  stellar masses are estimated from $K$-- and $R$--band luminosities.}
\end{center}
\end{table}

Based on our wavelet-based 0.4--5~keV point source search, Galaxy~A is
not detected in the {\em Chandra} data, possibly due to its proximity
to the central X-ray peak, whereas Galaxy~B is clearly detected. This
is true also if only considering a soft band (0.3--2~keV) source
search, whereas none of the galaxies is detected at higher energies
such as 2--8~keV. Galaxy~B is not robustly detected in 2--8~keV {\em
XMM} data either. All this immediately argues against significant
moderately obscured AGN activity in either galaxy, and potentially
also against the presence of a significant hot gas component in
Galaxy~A.

For Galaxy~A, any X-ray surface brightness analysis is severely
hampered by its proximity to the peak of the ICM emission $<10''$ to
the north.  If employing a 5--$10''$ semi-annulus to the south as
representative of the local background around the source, it is
formally detected at $2.7\sigma$ above this background in
0.3--2~keV. This is below the $3\sigma$ significance threshold adopted
for the adaptive smoothing in Figure~\ref{fig,cxomosaic}, explaining
its apparent absence in that figure. It is not detected in 1--3~keV
however, implying a background-subtracted (1--3~keV)/ (0.5--1~keV)
hardness ratio of $<0.35$ at 90\% confidence. This in turn implies
$\Gamma > 3.9$ for a power-law spectrum subject to Galactic
absorption, or $T<0.8$~keV for a thermal plasma with
$Z=0.4$~Z$_\odot$. This suggests that any emission from this source
has properties inconsistent with those of a typical AGN spectrum, but
leaves open the possibility of some contribution from hot gas.  As
further confirmation of the lack of substantial nuclear X-ray
activity, we note that the 0.5--8~keV count rate within $3''$ implies
$L_{\rm X} < 2\times 10^{40}$~erg~s$^{-1}$ in this band for any
intrinsically unobscured $\Gamma \approx 1.7$ component.  Furthermore,
since $3''$ still corresponds to a physical scale of $r\approx
3.5$~kpc, this upper limit contains a contribution from non-nuclear
emission.  We also note that we do not detect any point sources in the
stellar plume of this galaxy, down to a limiting 0.5--2~keV flux of
$\sim 3\times 10^{-16}$~erg~cm$^{-2}$~s$^{-1}$ (for a $\Gamma=1.7$
spectrum). This implies that there are no clusters of X-ray binaries
in the plume with a combined luminosity exceeding $L_{\rm X} \approx
2\times 10^{39}$~erg~s$^{-1}$, thus also ruling out the presence of
any ultra-luminous X-ray sources in this region.

In contrast to the case of Galaxy~A, there is a clear indication that
the soft emission associated with the optically fainter Galaxy~B is
spatially extended in the {\em Chandra} data.  Figure~\ref{fig,galB}
shows a 0.3--2~keV surface brightness profile extracted from the
optical galaxy center in bins of S/N $\ge 3$.  Emission is detected
out to $r=15''$ at $3\sigma$ above the local background, as evaluated
in a surrounding $25''$--$45''$ annulus. Both source and background
regions excluded position angles of $\pm 70^\circ$ around the North,
to suppress potential contamination from Galaxy~A and from local
variations in ICM emission.  The inferred profile is clearly much
broader than that of the $E=0.9$~keV {\em Chandra} point spread
function (PSF) at this detector position as estimated using the {\em
mkpsf} tool in {\sc ciao}. A Kolmogorov--Smirnov test yields a
probability of only $4\times 10^{-4}$ that the two data sets have been
drawn from the same distribution, and the source profile is well
described by a $\beta$--model with $\beta \approx 0.45$ as is typical
for extended emission from early-type galaxies (e.g.,
\citealt{osul03}).

\begin{figure}
\begin{center}
\epsscale{1.0} 
\plotone{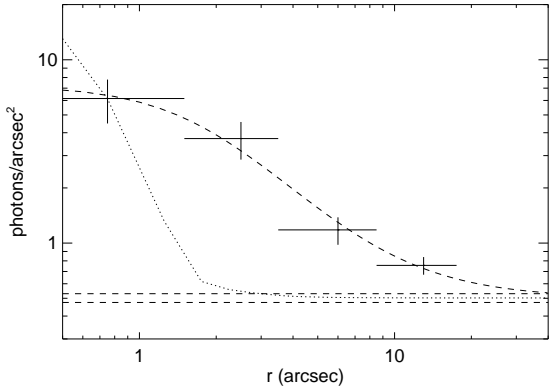}
\end{center}
\figcaption{Surface brightness profile of emission associated with
  Galaxy~B in bins of S/N~$\geq 3$. Dashed line shows the best-fit
  $\beta$--model, with $r_c=2\pm 1''$ and
  $\beta=0.45^{+0.10}_{-0.07}$. Dotted line shows the profile of the
  ACIS PSF at this position, normalized to match the innermost source
  data point.  Horizontal lines outline the 1-$\sigma$ uncertainties
  on the local background level.
\label{fig,galB}}
\end{figure}

At 95\% confidence, a simple absorbed power-law fit to the spectrum of
Galaxy~B within $r=15''$ would require $\Gamma > 3.8$ (for spectral
bins of 15 net counts, assuming standard $\chi^2$~statistics) or
$\Gamma > 2.3$ (assuming Cash statistics with bins of 5 net
counts). In addition, the fit shows obvious positive residuals at
$E\approx 1$~keV, strongly suggestive of an Fe peak. These results are
inconsistent with the expectation for emission dominated by an AGN or
low-mass X-ray binaries (which both tend to show $\Gamma \sim 1.7$),
but suggest an important contribution from thermal gas. An APEC model
with Galactic absorption returns a best-fit $T\approx 0.6\pm 0.2$~keV
for $Z$ fixed at 0.5~Z$_\odot$, with $L_{\rm X} \approx 3\times
10^{40}$~erg~s$^{-1}$ (0.3--2~keV). For these parameters, the observed
flux would imply a total hot gas mass of $\sim 1\times 10^9$~M$_\odot$
within $r=15''$ ($r\approx 17$~kpc).

Within $3''$ of the optical center, the (2--8~keV)/ (0.5--2~keV)
particle-subtracted hardness ratio of $0.07^{+0.05}_{-0.03}$ also
suggests rather soft emission. This result is difficult to reconcile
with the presence of an intrinsically bright but moderately obscured
source, since any intrinsically obscured power-law source with $\Gamma
<2$ should display a ratio of $>0.25$. With Galactic absorption only,
the observed ratio would instead suggest $\Gamma \approx 3.1$ ($>2.5$
at 1-$\sigma$), inconsistent with a typical AGN spectrum. The total
0.5--8~keV count rate within $r=3''$ happens to be identical to that
of Galaxy~A, again implying $L_{\rm X} < 2\times 10^{40}$~erg~s$^{-1}$
for any intrinsically unobscured $\Gamma=1.7$ AGN. To escape {\em
Chandra} detection in the 2--8~keV band, any AGN intrinsically
brighter than $10^{40}$~erg~s$^{-1}$ with a $\Gamma=1.7$ spectrum must
be subject to an intrinsic absorbing column in excess of $5\times
10^{23}$~cm$^{-2}$, and the source would have to be Compton-thick
($N_{\rm H}\ga 1.5\times 10^{24}$~cm$^{-2}$) for all intrinsic
luminosities $L_{\rm X} \ga 10^{42}$~erg$^{-1}$.

In summary, Galaxy~A which remains undetected at $24\,\mu$m shows
little evidence for a substantial AGN component, unless sufficiently
obscured to evade detection even at energies up to 8~keV. Even for the
IR-detected Galaxy~B, there is also no evidence for prominent nuclear
X-ray activity below this energy. The lack of significant hard X-ray
emission within the {\em Chandra} energy range in either object is
also supported by the hardness ratio map, in which both galaxies
appear softer than typical point sources in the field (some of which
are likely associated with background AGN). Regarding any hot ISM
material, tentative evidence for the presence of hot gas is found in
Galaxy~A, but we caution against strong conclusions in this case,
given that this galaxy is superposed on very bright ICM
emission. However, Galaxy~B provides clear evidence for extended
emission probably associated with hot ``halo'' gas. For its $B_J$--$R$
color of 2.1 derived above, even extreme estimates of the discrete
source contribution to the total $L_{\rm X}$ in early-types of the
resulting $L_B$ would suggest that $>80$\% of its X-ray output is due
to hot gas \citep{osul01}. Furthermore, its $L_{\rm X}$ within the
$15''$ radius of detection would place it above, but within the
scatter of, the relation between $L_B$ and $L_{\rm X}$ for local
early-types \citep{osul01}. The same applies when comparing it to the
expectation from the relation between $L_K$ and thermal $L_{\rm X}$
for early-types in groups as derived by \citet{jelt08}. These results
are all consistent with the idea that Galaxy~B contains significant
amounts of hot diffuse gas.

\section{DISCUSSION}\label{sec,discuss}

The ICM X-ray luminosity and emission-weighted mean temperature
derived from the {\em XMM} data, as listed in Table~\ref{tab,x_opt},
would place MZ\,10451 very close to the $L_{\rm X}$--$T$ relation for
X-ray bright groups of \citet{osmo04}. Combined with the regular X-ray
morphology on both small and large scales, and the clear evidence for
a cool core and a central metallicity excess seen in
Figures~\ref{fig,profiles} and \ref{fig,maps}, these features suggest a
system with ICM properties very typical of those of relaxed X-ray
bright groups and clusters. 

Before addressing the impact of the central galaxy merger on the ICM,
as well as the nature of the merger itself, it is useful to first
assess the current stage of the interaction. It is not immediately
clear from the present data whether the two galaxies have already
experienced a first close passage. The pronounced tidal stream
protruding northwards from Galaxy~A might suggest so, but dedicated
$N$-body simulations of binary mergers (which we hope to employ in
future work) would be needed to firmly establish under which orbital
conditions this feature could arise. One argument {\em against} this
possibility, however, is the observation of an extended thermal
component associated with Galaxy~B. It seems unlikely that a galactic
hot gas halo would survive a direct passage of one galaxy through the
other, especially since Galaxy~B would then have approached from the
north and so also have passed through the dense ICM core.

From the available information we can provide a rough estimate of the
time-scale for nuclear coalescence. To this end, we apply the
calibration of \citet{kitz08} based on the Millennium Simulation,
\begin{equation}
  t_{\rm merge} = 0.9 \frac{r_p}{25
  \mbox{~kpc}}\left(\frac{M_\ast}{10^{11}\mbox{~M}_\odot}\right)^{-0.3}\left(1
  + \frac{z}{8}\right) \mbox{ Gyr},
\label{eq,tmerge}
\end{equation}
where $M_\ast$ is the stellar mass of the merger product and $r_p$ the
current projected distance between the progenitors. Using $M_\ast
\approx 2.9\times 10^{11}$~M$_\odot$ from the numbers in
Table~\ref{tab,gals} along with $r_p \approx 25$~kpc, one obtains
$t_{\rm merge}\sim 700$~Myr. Hence, the galaxies are likely still at
least half a Gyr from coalescence, a number that is probably uncertain
by at least a factor of two. Table~\ref{tab,time} summarizes, in
ascending order, this and other timescales discussed in the following.

\begin{table*}
\caption{Timescales discussed in Section~\ref{sec,discuss}\label{tab,time}}
\begin{center}
\begin{tabular}{rll}
  \tableline \hline
Timescale  & Description & See \\ \hline
$\sim 75$~Myr & 
   Timescale for dispersal of ICM pressure enhancement between Galaxy~A and B &
   Section~\ref{sec,impact}, Figure~\ref{fig,maps}(c) \\
$\ga 150$~Myr & 
   Age of southern hotspot if representing a weak shock front propagating from
   ICM core & Section~\ref{sec,merger}, Figure~\ref{fig,maps}(a)  \\
$\sim 0.6$~Gyr & 
   Central ICM cooling time & Section~\ref{sec,impact}, 
   Figure~\ref{fig,mass}(c) \\
$\la 0.7$~Gyr &  
   Timescale for generation of central ICM metal excess by stellar plume 
   of Galaxy~A & Section~\ref{sec,impact} \\
$\sim 0.7$~Gyr & 
   Timescale for nuclear coalescence of Galaxy~A and B & Equation~(\ref{eq,tmerge}) \\
$\sim 1.4$~Gyr & Dynamical (free-fall) timescale at $r_{200}$  & 
   Section~\ref{sec,merger} \\
$\ga 2$~Gyr    & 
   Time since last major group--group merger in MZ\,10451 from X-ray 
   isophotes & Section~\ref{sec,merger}; \citet{pool06} \\ 
$\sim 2.4$~Gyr   & ICM sound crossing time within $r_{200}$ & 
   Section~\ref{sec,merger} \\
$\ga 200$~Gyr  & Timescale for MZ\,10451 to evolve into a ``fossil group'' &
   Equation~(\ref{eq,dynfric}) \\ \tableline
\end{tabular}
\end{center}
\end{table*}

\subsection{Impact of the Merger on the Group Environment}\label{sec,impact}

The regular X-ray morphology of MZ\,10451, even in the central
regions, suggests that the ongoing galaxy interaction has not had any
major impact on the ICM so far. In particular, although the merger has
significantly disrupted the stellar content near the center of
MZ\,10451, the system seems to have retained a cool core at this
stage. The possible destruction of cool cores, or the prevention of
them forming in the first place, is still poorly understood and widely
debated in the literature (e.g., \citealt{burn08,lecc10}). Our results
suggest that central galaxy--galaxy mergers do not necessarily
completely destroy cool cores through ICM heating or mixing, or that
any such activity occurs closer to the final merger stages (when
perhaps quasar-like activity is triggered in the case of gas-{\em
rich} major mergers; cf.\ \citealt{hopk08}). In partial support of the
former possibility, we note that dynamical heating by galaxy motions
alone generally seem insufficient to significantly disrupt cooling in
cluster cores \citep{falt05}.

The situation in MZ\,10451 is somewhat different from that in typical
X-ray clusters where a fully formed BCG is already residing at the
cluster center and potentially re-heating its surroundings via
intermittent AGN activity. Nevertheless, the offset between Galaxy~A
and the X-ray peak is still small, $\sim 10$~kpc, and the central
cooling time of $\sim 0.6$~Gyr is comparable to that of other
cool-core systems at the relevant radii \citep{sand06}. This further
supports the idea that any ICM mixing must have been modest in the
very core of MZ\,10451. This is also borne out by the entropy
distribution; while enhanced central entropy compared to the case of
typical cool-core systems could suggest a recent heating or mixing
episode, we find the central entropy to be quite low, and close to the
$\sim 30$~keV~cm$^2$ below which strong H$\alpha$ and radio emission
is typically seen in central BCGs \citep{cava08}, presumably as a
result of ongoing star formation fueled by radiative cooling of ICM
material.

Despite the apparent absence of strong ICM mixing, some features of
the ICM thermodynamic properties may be linked to the ongoing
interaction. Figure~\ref{fig,maps}(c) indicates the presence of a
region of enhanced ICM pressure between the two galaxies, with no
obvious counterpart in the surface brightness or entropy maps. In the
absence of external influences, this feature should disappear roughly
on its sound crossing timescale, which is $\sim 75$~Myr for its
estimated spatial extent of $\sim 0.5' \sim 35$~kpc. It is therefore
likely to be associated with a very recent or ongoing event, plausibly
reflecting ongoing adiabatic compression and heating of the ICM
between the galaxies, induced by their relative motion.

It remains unclear, however, whether the tentative presence of an
extended region of hot, high-entropy gas immediately south of Galaxy~B
is related to the interaction. The hardness ratio and entropy maps
suggest that the extent of the structure is several tens of kpc even
at its narrowest range. Thus, an explanation invoking a shock front
generated by Galaxy~B moving supersonically southwards seems excluded.
Alternatively, this gas could represent high-entropy material ejected
or stripped from either galaxy as a result of the interaction, but
again, the sheer size of this feature makes this unlikely. There are
insufficient counts in this region to test for departures from solar
abundance ratios, thus precluding any direct tests for a possible
origin in starburst outflows. However, such an explanation would be at
odds with the lack of clear post-starburst features in the optical
spectrum of either galaxy. It is possible that the feature is due to
``sloshing'' of the core gas owing to the gravitational perturbation
to the group potential caused by the ongoing merger, although the
absence of a similar feature in the surface brightness distribution
may argue against this \citep{mark01}. Another possibility, discussed
in more detail in the next Section, is that it is related to a
larger-scale interaction between two sub-groups.

From the viewpoint of the ICM abundance distribution,
Figure~\ref{fig,profiles} shows that any merger--induced ICM
enrichment is so far confined to the innermost $r\sim 0.5' \sim
30$~kpc and so has had no large-scale impact on the ICM metal content.
Nevertheless, the central abundance peak seen in this and many other
systems could be partly related to processes associated with BCG
formation. The observation of enhanced ICM abundances in the region
covered by the stellar plume of Galaxy~A may signify the presence of
stripped, highly enriched ISM from this galaxy. This would suggest
that central metal excesses could generally have a contribution from
enriched gas shed by central BCG progenitors. In fact, the ICM mixing
expected during such a merger may help to explain the considerable
{\em extent} of the central abundance peak in many systems, thus
alleviating the need for additional subsequent processes, such as
AGN--driven turbulence \citep{rebu05}, to disperse metals well beyond
the optical extent of the central BCG.

An alternative explanation would associate the central abundance peak
in MZ\,10451 with {\em in situ} enrichment by stars tidally stripped
during the interaction, some of which are likely to evolve into an
intracluster light component. To assess the feasibility of this
scenario, we first reiterate that the subsolar Si/Fe ratio seen in the
central region implies an important contribution from SN~Ia and hence
from old stars, in line with results for fully formed BCGs in X-ray
bright groups \citep{rasm09}. This also indicates that central ICM
enrichment from any young stellar population in the plume (or within
the possibly star-forming Galaxy~B) must be modest, in line with the
apparent absence of the plume in our UV and $24\,\mu$m images in
Figure~\ref{fig,galpair}. At the resolution of
Figure~\ref{fig,profiles}, the elevated ICM abundance in the group
core implies an {\em excess} Fe mass in the core of $\sim 1\times
10^6$~M$_\odot$.  Aperture photometry based on our Magellan $R$-band
image further suggests that $\sim 1/3$ of the stellar light of
Galaxy~A is located in this plume, indicating a plume stellar mass of
$\sim 6 \times 10^{10}$~M$_\odot$ for $M_\ast/L_R \approx
5.3$~M$_\odot/$L$_\odot$ as assumed in Section~\ref{sec,pair}. Given
these values, SN~Ia in the plume could have provided the central Fe
excess on a timescale of just $\sim 0.7$~Gyr, estimated using the
approach of \citet{rasm09} and excluding any contribution from SN~II
or stellar winds. Thus, it seems entirely feasible for intracluster
light generated by the merger to have produced the central Fe
excess. Forthcoming {\em Hubble Space Telescope} ({\em HST}) imaging
of MZ\,10451 will be used to provide more reliable mass and age
estimates of the stripped stellar component of Galaxy~A and further
test the viability of this picture.

\subsection{A Group--Group Merger? Gas and Galaxy Dynamics in MZ\,10451}\label{sec,merger}

The relatively undisturbed appearance of the ICM suggests that the
central BCG merger does not result as a consequence of a recent
larger-scale merger between separate subgroups. Idealized
hydrodynamical simulations of major cluster--cluster mergers with a
range of progenitor mass ratios and impact parameters suggest that the
timescale for a merged system to appear relaxed, as judged visually
from the morphology of its X-ray isophotes, is $\ga 2$~Gyr from the
time at which the gas core of the secondary merges with that of the
primary \citep{pool06}. For MZ\,10451 specifically, we note that the
ICM appears regular on the scales of Figure~\ref{fig,xmmmosaic}, i.e.\
out to at least $r \approx 300$~kpc. The ICM should relax roughly on a
sound crossing time $t_{\rm cr}$. Using Equation~(\ref{eq,tprof}) to
evaluate the local sound speed at all $r$ gives $t_{\rm cr} \sim
1.5$~Gyr within $r=300$~kpc, and $t_{\rm cr} \sim 2.4$~Gyr within
$r_{200} \approx 500$~kpc, in broad agreement with the general
expectation from the above simulations. This confirms that any major
group--group merger in MZ\,10451 cannot have taken place very
recently.

However, it may still be premature to entirely dismiss the possibility
of such a merger, partly because the observed entropy enhancement to
the south of the group core is not easily explained by a
galaxy--galaxy interaction alone, and partly because the ICM may have
relaxed more rapidly than the galaxies. We note that the entropy
feature is not seen in the surface brightness distribution of
Figure~\ref{fig,cxomosaic}, is not particularly sharply defined in the
entropy and temperature maps of Figure~\ref{fig,maps}, and has higher
temperature and entropy than its surroundings. It is therefore
unlikely to represent a cold front \citep{mark01} resulting from a
subgroup penetrating the group core, or a remnant cool core of such a
subgroup.

Nevertheless, it is still conceivable that the feature represents a
weak, merger-generated shock front, possibly viewed at some oblique
angle. If so, a lower limit to its age of $\sim 150$~Myr can be
estimated from its current projected distance to the X-ray core of
$\sim 70$~kpc, assuming propagation from the core at a velocity
comparable to the local sound speed. One might have expected a similar
feature to the north of the group core, but projection effects could
be rendering this undetectable. In addition, simulations of
cluster--cluster mergers \citep{pool06,mcca07} suggest that such
central entropy asymmetries, including the curved appearance of the
region in Figure~\ref{fig,maps}, {\em can} arise as a consequence of
shock heating in cluster mergers when the progenitor mass ratio
differs substantially from unity ($\ga$\,3:1). We note that
dissipation of turbulent energy injected by such a shock could easily
account for the observed entropy enhancement. The mean ICM density at
the approximate radial distance of this feature is $n \approx 1\times
10^{-3}$~cm$^{-3}$; for a cylindrical volume of length $\sim 1.2'$ and
base diameter $\sim 0.5'$ (assuming a depth along the line of sight
equal to the ``width'' of the feature), the total gas mass is then
$\sim 1\times 10^9$~M$_\odot$ within this region. The energy required
to raise $T$ of this gas by $\approx 0.15$~keV, as suggested by our
spectral analysis, is $\sim 1\times 10^{57}$~erg. This is three orders
of magnitude below what can be provided by shocks generated by a major
group--group merger producing a system of the relevant virial mass
\citep{paul10}, so energetically it seems entirely feasible to produce
this feature through shocks associated with an infalling subgroup.

Hence, the observed north--south entropy asymmetry could conceivably
result from the aftermath of a group--group merger, with Galaxy~A and
B representing the central galaxies of the two progenitor groups. It
is, in fact, possible that Galaxy~A was initially at rest at the X-ray
peak of the more massive subgroup but has now been slightly displaced
as a result of the final central galaxy--galaxy interaction. There are
arguments both for and against this interpretation. On the one hand,
the estimated time since the last major merger in MZ\,10451 of $\ga
2$~Gyr, as discussed above, is larger than both the dynamical
(free-fall) timescale at $r_{200}$ of $\sim 1.4$~Gyr and the current
estimated timescale for nuclear coalescence between Galaxy A and B
(cf.\ Table~\ref{tab,time}).  Hence, if each galaxy represented the
central galaxy of a progenitor system, it is perhaps surprising that
they have not yet themselves merged. On the other hand, we note that
the galaxy velocity distribution in MZ\,10451 does appear bimodal,
with the velocity histogram of the group members, shown in
Figure~\ref{fig,galdistr}(a), revealing a smaller, receding peak
relative to the system mean (containing 12 of the 60 confirmed group
members; \citealt{bai10}).

\begin{figure*}
\begin{center}
\epsscale{1} 
\plotone{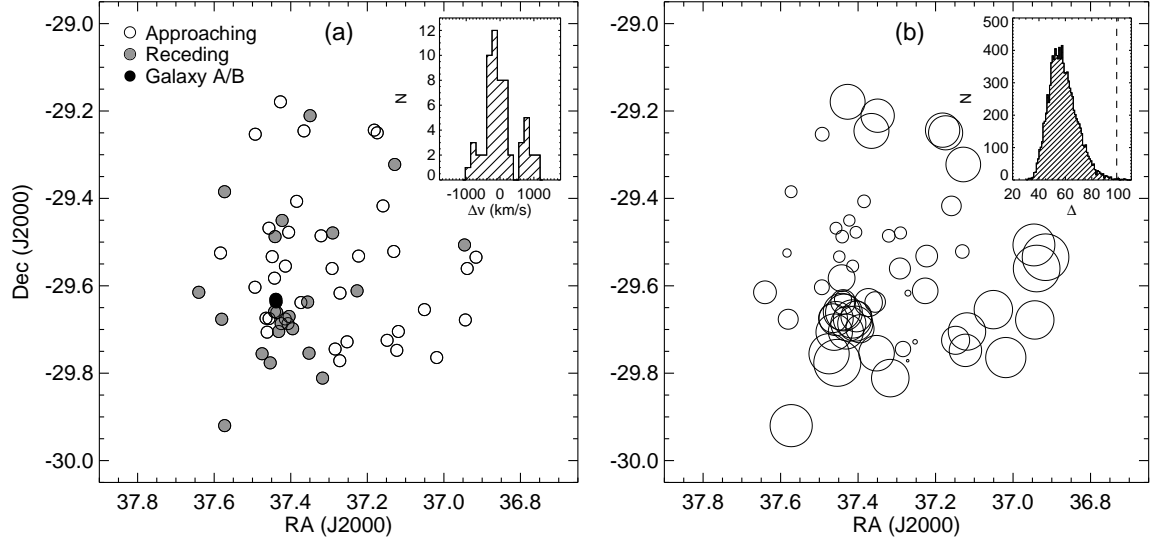}
\end{center}
\figcaption{(a) Spatial distribution of member galaxies in MZ\,10451,
  divided according to whether their redshift is smaller (white
  circles) or larger (gray circles) than the system mean. The merging
  central galaxy pair, both with redshifts consistent with the group
  mean, are shown by black circles. Inset shows the velocity histogram
  of the member galaxies. (b) Results of a Dressler--Shectman
  test. Each galaxy is marked by a circle whose diameter scales with
  the deviation $\delta_i$ of the local kinematics. Many large
  overlapping circles in an area indicate a correlated spatial and
  kinematic variation. Inset shows the distribution of $\Delta=\sum \delta_i$
  resulting from Monte Carlo calculations assuming purely random
  kinematic substructure, with the observed value for MZ\,10451,
  $\Delta=98.9$, marked by a dashed line.
\label{fig,galdistr}}
\end{figure*}

As also illustrated in Figure~\ref{fig,galdistr}(a), the spatial
distribution of galaxies in the group provides no clear indication
that infall of a subgroup, let alone a group--group major merger, has
recently taken place. To test this possibility in more detail and
search for evidence of kinematic substructure in the system, we
performed a Dressler--Shectman test \citep{dres88}. For each galaxy
$i$ with radial velocity $v_i$, we recorded the local kinematic
deviation $\delta_i$ from the global mean velocity $\langle v\rangle$
and velocity dispersion $\sigma_r$,
\begin{equation}
\delta_i^2 = (11/\sigma_r^2) [ (\langle v_{\rm local}\rangle -\langle
v\rangle)^2 + (\sigma_{\rm local} -\sigma_r)^2],
\end{equation}
using $\langle v_{\rm local}\rangle$ and $\sigma_{\rm local}$ derived
including its 10 nearest neighbors. We also recorded the cumulative
deviation $\Delta$,
\begin{equation}
\Delta = \sum_{i=1}^{N_{\rm gal}}  \delta_i ,
\end{equation} 
calibrating this result against 10,000 Monte Carlo simulations in
which all galaxy velocities $v_i$ were randomly shuffled. The latter
step provides an estimate of the probability $P$ of obtaining the
observed galaxy position--velocity configuration in the presence of
purely random substructure. We find $P \sim 0.2$\%, regardless of
whether all 60 members or only the 19 covered by the detectable X-ray
emission are included in the calculation. The results for $N_{\rm
gal}=60$ are illustrated in Figure~\ref{fig,galdistr}(b) and suggest
the presence of some non-random kinematic substructure in MZ\,10451,
both globally and in the central X-ray bright regions. 

Thus, the observed galaxy kinematics seem consistent with the idea
that infall of a smaller subgroup (possibly hosting Galaxy~B as the
central galaxy) has taken place. We also note that the estimated
timescales for several of the features in Figures~\ref{fig,profiles}
and \ref{fig,maps} to arise or survive (the cool core, central metal
excess, southern hotspot, and pressure enhancement between Galaxy~A
and B) are all consistent with being lower than the likely time since
such a merger (Table~\ref{tab,time}), so their presence seems
compatible with this possibility. However, we stress that this
interpretation must be regarded as tentative, is not without its
problems, and that more robust tests of the group--group merger
scenario would require detailed spectral mapping of the group core and
hence deeper X-ray observations.

\subsection{The Nature of the Galaxy Merger: Implications for BCG Formation}\label{sec,implic}

In the $R$-band, the two central galaxies show a luminosity ratio of
2:1, with the brighter Galaxy~A having a luminosity comparable to an
$L_\ast$ galaxy \citep{zabl00}. They currently represent the first and
fifth-ranked group members within the derived virial radius of
$r_{100} \approx 0.7$~Mpc, although brighter galaxies with concordant
redshifts are present within $2r_{\rm vir}$ from the X-ray
peak. However, the combined $R$-band luminosity of the central pair,
$L_R \approx 5.4 \times10^{10}$~L$_\odot$, is $\sim 60$\% larger than
that of any of those galaxies within this radius. Hence, the merger
will comfortably produce the brightest galaxy within the group, thus
confirming it as a BCG major merger.

Recent {\em Chandra} observations show that most luminous early-type
galaxies maintain thermal X-ray halos even near the centers of groups
and rich clusters \citep{sun07,jelt08}. For example, excluding central
BCGs, \citet{jelt08} found that $\sim 80$\% of early-type galaxies in
groups with $L_K > L_{K,\ast} = 1.2\times 10^{11}$~L$_\odot$ retain a
hot gas halo. Both our galaxies have $L_K$ comparable to this limit
(Section~\ref{sec,pair}). The presence of extended thermal emission
associated with Galaxy~B indicates that it, too, possesses a hot halo,
suggesting that such halos may be preserved at least at this merger
stage. In the case of Galaxy~A, no such halo is unambiguously
detected, so it remains a possibility that any extended halo of this
galaxy has been largely disrupted, since this galaxy also appears
particularly disturbed in the optical. We should note, however, that
we do not have the S/N or spatial resolution to directly establish
whether either galaxy shows current evidence for hot gas stripping,
although the pseudo-entropy map (Figure~\ref{fig,maps}) suggests that
any such activity must be modest.

It is commonly assumed, as also suggested by numerical simulations,
that the nature of low-redshift major mergers leading to bright,
early-type BCGs is largely dissipationless, involving progenitors
containing little or no gas (e.g., \citealt{khoc03,boyl06,delu07}).
Based on observed stellar rotational velocities in nearby BCGs, some
likely counterexamples do exist however, although these seem rather
rare \citep{loub08}. It is therefore of interest to establish to what
extent either of the merging galaxies in MZ\,10451 is experiencing
active star formation fueled by a reservoir of {\em cold} gas.

In the case of Galaxy~A, the observed NUV flux would suggest weak star
formation at a level of $\sim 0.2$~M$_\odot$~yr$^{-1}$, but no IR
emission is accompanying the UV output down to our approximate
detection limit of 0.1~M$_\odot$~yr$^{-1}$ \citep{bai10}. In addition,
its FUV--NUV color of $\ga 1$ is consistent with that seen for the
majority of passively evolving ellipticals \citep{gild07}, suggesting
that the UV light may be dominated by that of an evolved stellar
population. 

While this latter point remains true also for Galaxy~B, this case is
nevertheless more ambiguous. The galaxy is detected at 24\,$\mu$m, but
there is no clear evidence for optical emission lines or nuclear X-ray
activity. Any AGN responsible for the 24\,$\mu$m emission must thus be
subject to strong intrinsic obscuration, with an absorbing column
exceeding $10^{24}$~cm$^{-2}$ for typical AGN X-ray luminosities.
Examples of such highly obscured AGN do occur at low redshift (see,
e.g., \citealt{coma04}), including, as is the case for Galaxy~B,
galaxies that do not show any Seyfert signatures in the optical
band. Alternatively, the infrared emission may instead be powered by
dusty star formation, implying that the galaxy also retains some cold
gas. If so, the observed 24\,$\mu$m flux would suggest low-level star
formation at a rate of $\sim 0.2$~M$_\odot$~yr$^{-1}$, computed using
the prescription in \citet{bai10} which takes into account the
contribution from cold dust heated by an evolved stellar
population. If attributed to young stars, the {\em GALEX} NUV flux
would instead suggest a rate of $\sim 0.05$~M$_\odot$~yr$^{-1}$ based
on the \citet{kenn98} relation and before correction for dust
attenuation. This is consistent with the idea that some obscured star
formation activity is taking place, supporting the notion of a cold
gas reservoir in Galaxy~B.  Unfortunately, the absence of a robust
{\em Spitzer} detection of either galaxy at 70\,$\mu$m implies that we
cannot use the 24/70\,$\mu$m flux ratio to distinguish between star
formation and AGN activity. Furthermore, the available NVSS radio data
of the galaxy pair are not deep enough to provide useful constraints
in this regard. The radio luminosity limits are well below a commonly
adopted dividing line between starburst galaxies and radio-loud AGN of
$L_{\rm 1.4\,GHz} \sim 3\times 10^{23}$~W~Hz$^{-1}$ (e.g.,
\citealt{yun01}), but while the limit on the $q_{24}$ parameter for
Galaxy~B, $q_{24} = \mbox{log}\,(S_{24\,\mu\mbox{m}}/S_{\rm 1.4\,GHz})
> -0.55$, is consistent with that of star-forming galaxies (e.g.,
\citealt{appl04}), it does not rule out a contribution from AGN
activity to the 24\,$\mu$m emission.

Deep H{\sc i} or CO observations would thus be needed to decisively
test for the presence of a cold gas reservoir in Galaxy~B. While such
data will eventually be provided by the upcoming WALLABY
survey\footnote{http://www.atnf.csiro.au/people/bkoribal/askap/} to be
undertaken by the Australian SKA Pathfinder array, we simply note here
that the tentative evidence for obscured star formation within
Galaxy~B points to an early-type major merger which does involve some
gas, with the implication that at least some low-redshift BCG major
mergers must be dissipational to some degree. Coupled with the absence
of pronounced nuclear activity in either galaxy, this result is
consistent with a picture in which the early stages of interactions
between early-type galaxies first trigger residual star formation,
followed by efficient central inflow of gas leading to strong AGN
activity at the final stages (e.g., \citealt{roge09}).

If identifying the peak of diffuse X-ray emission with the center of
the group gravitational potential, our results also suggest that BCGs
may form very close to the latter. This is consistent with a picture
in which central BCGs have been built up through a series of mergers
near cluster cores, in line with the galactic cannibalism scenario for
BCG growth discussed by \citet{ostr77}. As mentioned, it is possible
that Galaxy~A was initially at rest at the position of the X-ray peak
but has now been slightly displaced as a result of the ongoing
interaction. Furthermore, our radial velocity measurements and the
fact that the merging pair and the X-ray peak are all aligned,
indicate that the motion of the galaxies occurs largely in radial
orbits along a north-south axis, coinciding with the direction defined
by the stellar plume of Galaxy~A. At the same time, the large-scale
X-ray morphology seen in Figure~\ref{fig,xmmmosaic} implies an overall
ICM elongation along a similar axis. If the elongation of the ICM
betrays the orientation of the filaments feeding the cluster, then the
merging galaxies and their "parent" filaments line up. This is
consistent with expectations from simulations, which suggest that the
mergers forming central BCGs occur preferentially along radial orbits
aligned with the filaments feeding the host cluster, and thus with the
major axis of the cluster itself (\citealt{west94, boyl06} and
references therein).

\subsection{Properties of the Merger Product}

As a final point, it is interesting to also consider the likely
properties of the merger product, in terms of the nature of the merged
central galaxy and of the group as a whole. As mentioned, the
estimated $K$-band luminosity of the merged galaxy would be $L_{K,{\rm
BCG}} \approx 2.9\times 10^{11}$~L$_\odot$. Using $M_{200}$ from
Table~\ref{tab,x_opt}, we can compare this to the expectation based on
the observed low-redshift correlations between BCG $K$-band luminosity
and $M_{200}$ of the host cluster derived by \citet{lin04} and
\citet{brou08}. Either of these relations would suggest $L_K \sim
3\times 10^{11}$~L$_\odot$ for a BCG in a group of this mass, in
excellent agreement with the above estimate. While some of the stellar
mass of the BCG progenitors will be tidally stripped in the merging
process and thus will not contribute directly to $L_{K,{\rm BCG}}$
\citep{stan06}, it seems likely that MZ\,10451 in the post-merger
phase would thus have properties in good agreement with the observed
$L_{\rm K,BCG}$--$M_{200}$ relations, particularly in light of the
very considerable scatter around these relations at low values of
$M_{200}$.

The tidally stripped stars seen around both merging galaxies may
become unbound during the interaction and would then contribute to the
emergence of an intracluster light component. This possibility will be
investigated in more detail in a future paper based on our forthcoming
{\em HST} imaging of this system. However, we note for now that our
existing imaging of MZ\,10451 seems consistent with the notion that an
intracluster light component can form in systems which do not already
have a central dominant early-type galaxy.  Such a scenario draws
support from cosmological simulations which suggest that major mergers
associated with BCG formation play a prominent, and possibly dominant,
role in the generation of intracluster light \citep{mura07}.

It is also interesting to note that MZ\,10451 would not become a
``fossil group'' as a consequence of the merger, if adopting the
definition of \citet{jone03}. Although the X-ray luminosity criterion,
$L_{\rm X,bol} \geq 10^{42} h_{50}^{-2}$~erg~s$^{-1}$, is already
safely met, the combined magnitude of the merger product of $m_R
\approx 14.70$ would be only $\approx 1.2$~mag brighter than that of
the subsequently second-brightest galaxy within $0.5r_{\rm vir}
\approx 0.5r_{100} \approx 330$~kpc, thus falling short of the $\Delta
m_{12}=2$ gap between the first- and second-ranked galaxy required to
meet the above definition of a fossil system. This remains true even
if just considering the region within $0.5r_{200}\approx 250$~kpc, for
which $\Delta m_{12}$ would be $\approx 1.7$~mag.

Taking this argument even further, we note that any subsequent merger
between the forming central galaxy and the resulting brightest
satellite galaxy within a projected distance of $0.5r_{\rm vir}$ (with
$L_R \approx 1.2\times 10^{10}$~M$_\odot$) {\em would} create an
$R$-band magnitude gap of $\approx 2$ ($\Delta m_{12}=1.97$).  This
satellite currently has a radial velocity $\Delta v_r =
560$~km~s$^{-1}$ relative to the group mean and so is likely to be in
a low-eccentricity orbit. Under these circumstances, dynamical
friction acting on this galaxy could accomplish such a merger on a
time-scale of
\begin{equation}
  t_{\rm fric} \approx 12.4\, \frac{r_0}{100\,\mbox{kpc}}
  \left(\frac{v_H}{700\,\mbox{km~s$^{-1}$}} \right)^2
  \left(\frac{v_s}{250\,\mbox{km~s$^{-1}$}} \right)^{-3}\mbox{ Gyr},
\label{eq,dynfric}
\end{equation}
where $r_0$ and $v_H$ is the current orbital radius and circular
velocity of the satellite, respectively, and $v_s$ the characteristic
circular velocity of its dark matter halo \citep{dong05}.  A lower
limit to $r_0$ is the projected distance of the satellite from the
X-ray peak of $r_{\rm proj}=4.2' \approx 280$~kpc. Taking $v_H \sim
\Delta v_r$ and assuming an $R$-band total mass-to-light ratio of 10
for the satellite (a spiral), along with the assumption of an
isothermal sphere mass distribution for its halo, we have $v_s \sim
120$~km~s$^{-1}$.  With these numbers, equation~(\ref{eq,dynfric})
would predict a time-scale $t_{\rm fric} \ga 200$~Gyr, with equality
corresponding to $r_0 = r_{\rm proj}$.  Thus, even following the
ongoing central merger, MZ\,10451 would be resilient to the formation
of a fossil-like configuration by means of dynamical friction for many
Hubble times. This is consistent with the notion that present-day
fossil systems are unlikely to have formed recently (e.g.,
\citealt{khos07}).

\section{SUMMARY AND CONCLUSIONS}

The $z\approx 0.06$ low-mass cluster MZ\,10451 is, to our knowledge, a
unique system. It is X-ray bright, harbors a central BCG currently
being formed from the major merger of two optically luminous
early-type galaxies, and resides at a distance where the nature of the
BCG merger and its impact on the surrounding ICM can be studied in
unprecedented detail. We have combined {\em Chandra} observations of
the ICM in the cluster core with {\em XMM} data of the large-scale ICM
properties, and with optical, infrared, and UV data of the merging
galaxies, to explore the impact of the interaction on the ICM and on
the interacting galaxies themselves.

 The data indicate detectable ICM emission out to at least $r
 \approx 6'$, centered close to the two interacting galaxies. The ICM
 surface brightness in MZ\,10451 is symmetrically distributed on both
 small and large scales, consistent with the expectation for a
 virialized system. Hence, the ongoing central galaxy--galaxy
 interaction has not significantly disturbed the ICM density
 distribution even in the group core. On the largest scales, the ICM
 distribution is elongated in the north-south direction, however.

 The ICM exhibits a cool core and a central metal excess (here
 extending to $r\sim 30$~kpc), as typically seen in undisturbed X-ray
 bright systems with a central BCG.  This implies either that cool
 cores and central abundance peaks are not necessarily destroyed by
 central galaxy--galaxy mergers in clusters, or that any significant
 central ICM heating and mixing takes place at post-coalescence
 stages.

 The brighter of the merging galaxies presents a pronounced plume of
 tidally stripped stars. The ICM metallicity in the region covered by
 this feature is significantly enhanced. This suggests efficient local
 enrichment by intracluster stars ejected by the interaction, or
 merger-induced stripping of highly enriched gas from either
 galaxy. Such processes may thus have contributed to building up the
 pronounced central abundance peak seen in this and other systems.
 Some of the tidally stripped stars may represent the emergence of an
 intracluster light component, suggesting that such a component can
 form in systems that do not already host a central dominant galaxy.

 Both the merging galaxies are spatially offset from the ICM X-ray
 peak, but only by $\sim 10$ and $\sim 30$~kpc, and their radial
 velocities are both consistent with the group mean within the errors.
 This demonstrates that central BCG formation can occur close to the
 center of the cluster potential, consistent with the notion that
 central BCGs are built up through a series of mergers in cluster
 cores.

 The merger axis extends toward the X-ray peak and roughly coincides
 with the axis defined by the large-scale elongation of the
 ICM. Combined with the above, this provides support for a picture in
 which BCG mergers are caused by galaxies falling along radial orbits
 aligned with the accreting filaments that feed cluster growth.

 A rough estimate based on their current separation suggest that the
 interacting galaxies are still $\sim 0.5$~Gyr from nuclear
 coalescence. Despite being involved in a galaxy--galaxy interaction
 in the core of an X-ray bright system, at least one of the merging
 galaxies shows clear evidence of having retained an extended hot gas
 halo at this stage. The inferred infrared, UV, and X-ray properties
 of the galaxy pair are consistent with the presence of low-level
 obscured star formation in at least one of them. This would be
 consistent with a picture in which early-type galaxy interactions
 initially trigger residual star formation, followed by an
 AGN-dominated phase closer to coalescence. The tentative evidence for
 ongoing star formation in one of the merging galaxies suggests that
 low-redshift BCG major mergers need not be completely
 dissipationless. However, while the X-ray data reveal no evidence for
 strong moderately obscured AGN activity in either galaxy, the
 possibility of an AGN hidden behind a Compton-thick screen cannot be
 ruled out as an alternative to the presence of cold gas. Deeper radio
 observations would be required to decisively settle this issue.

Our forthcoming {\em HST} observations of MZ\,10451 should further
elucidate the optical properties of the central merging galaxies, and
thus help constrain the properties of the merger product itself. These
data will also help in establishing the age of tidally stripped stars,
and the generation and properties of any intracluster light in this
unusual system. A detailed analysis of these and our Magellan optical
data will be presented in a forthcoming paper.

\acknowledgments

We are very grateful to the referee for insightful and constructive
suggestions which improved the presentation of our results. This
research has made use of the NASA/IPAC Extragalactic Database
(NED). JR acknowledges support provided by the National Aeronautics
and Space Administration through Chandra Postdoctoral Fellowship Award
Number PF7-80050 issued by the Chandra X-ray Observatory Center, which
is operated by the Smithsonian Astrophysical Observatory for and on
behalf of the National Aeronautics and Space Administration under
contract NAS8-03060. JSM acknowledges partial support for this study
from Chandra grant GO9-0144X.

\end{document}